\documentclass{aa}  

\usepackage{graphicx}
\usepackage{txfonts}
\usepackage{multirow}
\usepackage[colorlinks=true,citecolor=blue,linkcolor=blue]{hyperref}
\usepackage[dvipsnames,namedcolors]{xcolor}
\usepackage{ulem}

\usepackage{etoolbox}

\begin{document}

   \title{Eclipse mapping study of the eclipsing binary KIC~3858884 with hybrid $\delta$~Sct/$\gamma$~Dor component} 
   \titlerunning{EM study of KIC~3858884}


   \author{A. B{\'o}kon
          \inst{1}\fnmsep\inst{2}\fnmsep\inst{3}\fnmsep\inst{4}\fnmsep\thanks{E-mail: andrasb@titan.physx.u-szeged.hu},
          I. B. B{\'i}r{\'o}\inst{5}\fnmsep\inst{6},
          A. Derekas\inst{2}\fnmsep\inst{4}\fnmsep\inst{6}
          }
    \authorrunning{A. B{\'o}kon et al.}

   \institute{Institute of Physics, Faculty of Sciences and Informatics, University of Szeged, D{\'o}m t{\'e}r 9, 6720, Szeged, Hungary
         \and
              ELTE E\"otv\"os L\'or\'and University, Gothard Astrophysical Observatory, Szombathely, Szent Imre h. u. 112, H-9700, Hungary,
         \and
              HUN-REN-ELTE Exoplanet Research Group, Szent Imre h. u. 112, 9700, Szombathely, Hungary,
         \and
              MTA-ELTE Lendület "Momentum" Milky Way Research Group, Szent Imre h. u. 112, 9700, Szombathely, Hungary
         \and
             Baja Astronomical Observatory of University of Szeged, 6500, Baja, Szegedi út, Kt. 766, Hungary
         \and
             HUN-REN-SZTE Stellar Astrophysics Research Group, 6500, Baja, Szegedi út, Kt. 766, Hungary
             }

   \date{Received ...; accepted ...}

 
  \abstract
   {}
   {Pulsating stars in eclipsing binary systems offer a unique possibility to empirically identify pulsation modes using the geometric effect of the eclipses on the pulsation signals. 
   Here we explore the $\delta$ Scuti type pulsations in the eclipsing binary system KIC~3858884 with the aim of identifying the dominant modes using various photometric methods.}
   {We used the \textit{Kepler} short cadence photometry data. Refined binary model and pulsation parameters were determined using an iterative separation of the eclipsing binary and pulsation signals.
   We used the photometric residuals, a phase modulation study, and a double eclipse mapping to identify the host stars of the dominant pulsations. 
   \'{E}chelle diagram diagnostics were employed to locate the frequencies with most influence from the eclipses.
   Direct Fitting methods assuming spherical harmonic surface patterns were explored 
   to determine an orientation for the symmetry axis and to infer surface mode numbers $\ell$ and $m$. General surface patterns were reconstructed using dynamic eclipse mapping, and provided ancillary mode number estimates.
   The use of these methods allowed us independent mode identification from asteroseismic models.
   }
   {We unambiguously established the secondary star as the main source of the pulsations. Seven peaks, including the two strongest modes, were found to show pronounced modulations during the secondary eclipses. For the first time ever, we could detect two hidden modes with amplitude intensification during the eclipses. Only one frequency appears to originate from the primary. 
   Direct Fitting yielded an essentially aligned pulsation axis for the secondary, which however requires validation by additional spectroscopic measurements. We successfully reconstructed surface patterns and determined mode numbers for most of the selected frequencies with both of our methods. One radial and three sectoral modes, (1,-1),  (2,$\pm$2) and (3,$\pm$3), were found. The two hidden modes were identified as (3,$\pm$1) and (2,$\pm$1). One additional radial mode turned out to be a combination frequency. Partial disagreement between EM and DF results may indicate that either the strongest modes deviate from strict spherical harmonics, or that the pulsation axis cannot be constrained by the photometric data alone. Future spectroscopic observations could help in resolving the question.}
   {}


   \maketitle
%

\section{Introduction}

Stellar oscillations may provide a unique tool for probing stellar interiors and determining their global properties like mass and radius.
Oscillations occur at various stages of stellar evolution, across the Hertzsprung-Russell Diagram. Therefore, performing this complex analysis holds great potential as a valuable probe for modern astrophysics.

A successful asteroseismic analysis requires the identification of the oscillation modes. This key step is easier for solar-like oscillations where pattern recognition in their observed oscillation frequency spectrum can be applied together with scaling relations \citep{kjeldsen.bedding.1995}. On the other hand, classical pulsators like the $\delta$~Sct and $\gamma$~Dor classes have complex, irregular frequency spectra, making the usual approach difficult to use.
Several mode identification methods have been developed for single stars, like exploiting the sensitivity of limb darkening to different pulsation modes with multi-colour photometry \citep{balona.evers.1999,Garrido_2000,dupret.etal.2003a,dupret.etal.2003b} or fitting the variations of the spectral line profiles with pulsation models \citep{aerts.eyer.2000,Briquet_2003}.
These methods rely heavily on detailed astrophysical models of stellar structure, atmosphere and pulsation, while spectroscopic methods require in addition high temporal and spectral resolution data, not easy to obtain; therefore these approaches have limited applicability.

A blooming era of asteroseismology began with the advent of the remarkably successful space-borne photometric missions MOST, CoRoT, \textit{Kepler} and TESS 
\citep{MOST2003PASP..115.1023W,CoRoT2009A&A...506..411A,Borucki2010Sci...327..977B,TESSRicker2015JATIS...1a4003R}, and will certainly continue with the commencing of similar future missions like the ESA PLATO mission, where asteroseismology will play a major role in determining stellar and planetary properties of the detected exoplanetary systems. Beside successfully completing their main objective of exoplanet observations, the extreme precision and high temporal resolution of the data pouring nearly continuously from these state of the art instruments revolutionised about every field of stellar studies. 
For pulsating stars the increased detectability of lower amplitudes and frequencies, coupled with unprecedented frequency resolution, furnished more complete frequency spectra, providing substantial benefits for asteroseismic studies.
It proved to be particularly useful for investigating solar-like oscillations \citep{chaplin.miglio.2013}, but notable 
progress has been achieved for classical pulsators, too (\citealp{garcia-hernandez.etal.2015} for $\delta$~Sct, in particular \citealp{bedding.etal.2020} for the young ones; \citealp{bedding.etal.2015,vanreeth.etal.2016,ouazzani.etal.2017,li.etal.2019}  for $\gamma$~Dor).

Another class of benchmark astrophysical objects are the binary and multiple systems, because they allow a precise determination of absolute parameters like stellar masses. Eclipsing systems are even more important because the study of the eclipses also yields radii, surface temperatures and distances. Eclipsing double-lined binary (EB+SB2) systems are the best cases because they basically yield all the above key parameters. In hierarchical triple eclipsing systems masses can be derived even without radial velocity measurements \citep{borkovits.etal.2016}. 
There are also an increasing number of binary and multiple systems being discovered to contain at least one pulsating component.
In this serendipitous configuration binarity can be used to determine absolute global parameters, a tremendous help for an asteroseismic investigation, which in turn yields internal structure and evolutionary status of the pulsating components.
The first systematic catalogue of pulsating stars in binaries was presented by \citet{szatmary.1990.catalog}. %
Starting from there, the number of such systems increased steadily, involving mostly classical pulsators like $\delta$~Scutis \citep{rodriguez.2000.catalog,rodriguez.breger.2001}. The space missions increased drastically not only the number of eclipsing binaries by discovering thousands of such systems, but also the number of pulsating members. 
The cataloue of \citet{zhou.2014.catalog} contained 515 pulsating binaries as of October 2014, more than half of them being eclipsing systems, with 96 oscillating Algol-type eclipsing binaries ('oEA', \citealp{mkrtichian2002ASPC..259...96M}). 
\citet{liakos.niarchos.2017} listed 199 $\delta$~Sct pulsators in binaries; \citet{gaulme.guzik.2019.catalog} identified 303 systems (163 new) from \textit{Kepler} light curves, containing $\delta$~Sct, $\gamma$~Dor, red giants and tidally excited oscillators. 
\citet{shi.2022.catalog} found 54 pulsators in EA type eclipsing binaries from TESS data, while
recently \citet{eze.handler.2024.catalog} reported on 78 $\beta$~Cep pulsators in eclipsing binaries from the same database.
(The potential lying in space-borne data is nicely illustrated e.g. by the results of \citet{zhou.2023.catalog} where, as a result of a targeted search, 1530 new variables have been found in a region with 1 degree radius of the open cluster NGC~6871, with almost 200 eclipsing binaries, 12 of them exhibiting pulsating or rotating components.)

Binaries containing pulsating stars also offer a unique possibility to probe astrodynamics from various perspectives, like the influence of mass transfer from the companion ('oEA' or oscillating eclipsing systems of Algol type, \citealp{mkrtichian2002ASPC..259...96M} or its tidal influence on the oscillations; this latter manifesting itself in a wide variety of effects, depending on the strength of the tidal forces: tidally perturbed oscillations \citep{reyniers.smeyers.2003a,reyniers.smeyers.2003b,Bowman2019ApJ...883L..26B,johnston.etal.2023}, tidally excited oscillations (the 'heartbeat' stars, \citealp{welsh2011ApJS..197....4W,guo.etal.2019});
tidal locking of the pulsation axis or even tidal trapping of the oscillations on the facing side of the star \citep{fuller.etal.2020,handler.etal.2020,shi.etal.2021}.

In addition there is also a purely visual effect of the eclipses on the observed pulsation signals, which can be used to infer their properties, potentially leading to the identification of the individual modes. Changes of the viewing geometry temporarily and temporally alter the visibility of the surface of the star undergoing the eclipses, changing the amount of light in the integrated flux {in addition to} the physical brightness changes on the surface due to the pulsations. This ultimately leads to 
a modulation of the pulsations during the eclipses, depending on the surface patterns. With apropriate modelling the nature of the surface patterns can be recovered from these modulations \citep{gamarova.etal.2003,reed.etal.2005,nuspl.etal.2004}
It was also recently exploited by \citet{johnston.etal.2023} to investigate the tidal influence of the pulsations in the eclipsing binary U~Gru.

It is interesting to identify modes in more detached systems with free oscillations intrinsic to the star and marginal tidal influence from its companion, because it has the potantial of applying classical asteroseismic inversions to the components.

KIC~3858884 is a detached eclipsing binary with a wide eccentric orbit of orbital period $\sim$27 days and eccentricity $e\sim 0,47$. It was first investigated by \cite{Maceroni.etal.2014}, whose work established the first comprehensive analysis of the system regarding the aspects of both an eclipsing binary model and of pulsations. Most of the pulsation signal seems to originate on the secondary, which is a $\delta$ Sct/$\gamma$ Dor hybrid pulsator, but the primary component is also suspected to show pulsations, because the two stars are very similar: mass ratio close to 1 ($q=0.987$), temperatures and radii also similar ($T_2/T_1 \sim 0.97$, $r_2/r_1 \sim 0.88$). \cite{Manzoori2020MNRAS.498.1871M} investigated the linear and nonlinear effects of tidal interaction on the combination frequencies possibly occurring in the system. %

We used the geometric modulating effect of the eclipses on the pulsation signals to identify individual pulsation modes by image reconstruction (eclipse mapping) as a general method and by an MCMC exploration of spherical harmonics as a more model-specific method. For the analysis, we used \textit{Kepler}  photometric data.
Section~\ref{sec:data} presents the data preparation and detrending followed by a description of the iterative process of disentangling the eclipse and pulsation components of the light curve in Section~\ref{sec:disentangling}. Section~\ref{sec:puls-source} describes the used methods for identifying the source stars originating the frequencies on. Section~\ref{sec:modeid} presents the mode identification procedures and the results. In Section~\ref{sec:discussion} we discuss the results and summarise our conclusions.


\section{Data acquisition and processing}
\label{sec:data}

KIC~3858884 was observed by \textit{Kepler} in the original Kepler field. Long cadence data (LC, $\sim30$~min effective exposure time) were collected in every quarter except Q6, Q10 and Q14. Short cadence (SC, exp time $\sim$2 min) observations were performed in Q2, Q8 and Q9. In our analysis, we used the short cadence data because the
characteristic period of the pulsations has
the order of minutes.
They cover 9 months of data series and approximately the same number of binary orbits.

\begin{figure}
\begin{center}
\includegraphics[width=\columnwidth]{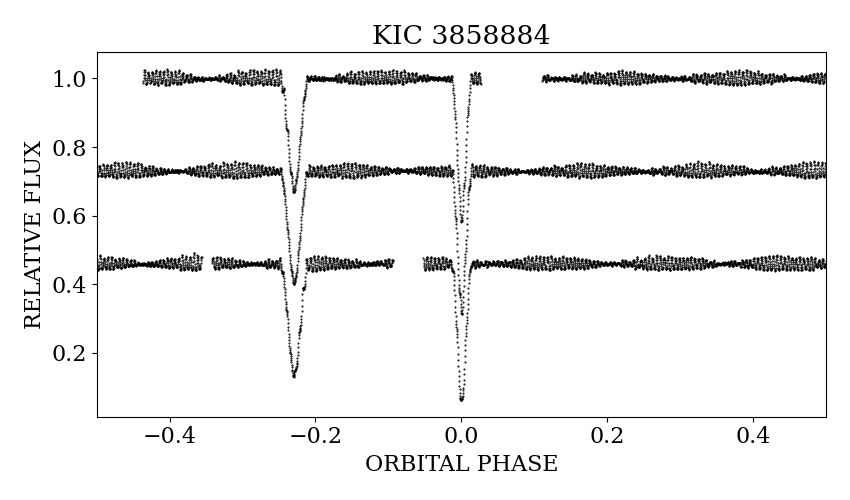}
\caption{Sample light curves of KIC~3858884 after detrending, folded within one orbit and shifted vertically by 0.27 from each other for illustration purposes. The first three observed orbits are shown. Only every 12th data point is drawn for {better visibility}.}
\label{fig::0_sample}
\end{center}
\end{figure}

For the analysis we downloaded the Simple Aperture Photometry (SAP) flux data from the  Mikulski Archive for Space Telescopes (MAST) database. 
The trends were determined by a careful choice of co-trending base vectors 
using tasks from the \texttt{pyKE}\footnote{https://nexsci.caltech.edu/workshop/2012/keplergo/PyKE.shtml} package. The remaining trends were removed by fitting a low order polynomial to the out of eclipse envelope of the light curve using a custom written code. The light curve segments were then normalised in flux to out of eclipse levels of 1. We completed these procedures separately for each quarter. A sample of the detrended light curve is shown on the top panel of Fig.~\ref{fig::0_sample}.

\section{Disentangling the light curve}
\label{sec:disentangling}

Due to the strong interdependence of the eclipse and pulsation characteristics there is no reliable method yet to simultaneously fit the eclipses and the pulsations on the data. 
Although it would be tempting to use e.g. wavelet-based algorithms to automatically fit the pulsations during a binary modelling as red noise or Gaussian processes and remove them in one step, we did not want to delegate the delicate task of disentangling to a task that has no notion of eclipse-modulated pulsations. Besides, such algorithms would probably have difficulty dealing with hundreds of simultaneous pulsations.
Rather, we carried out a traditional iterative procedure of separating the eclipsing binary signal (the light contributions of the equilibrium surface brightness maps of the stars) and the pulsation signals.
The methods analysing eclipse-modulated pulsations operate on the pulsation-only signal, while of course also relying on the binary model, to be determined from the binary signal.

\begin{figure}
\begin{center}
\includegraphics[width=\columnwidth]{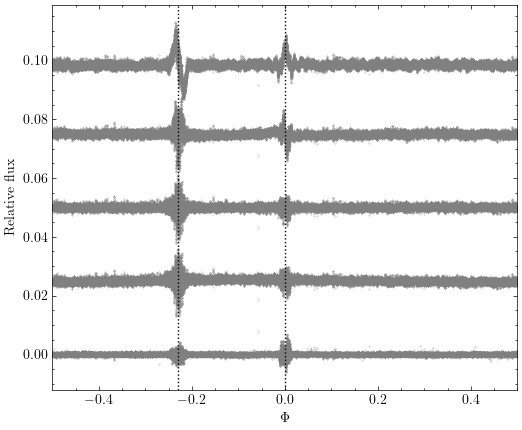}
\end{center}
\caption{The evolution of the residuals from the light curve disentangling process for KIC~3858884 across the five iterations (from top to bottom). All points from the full data set are plotted. The centers of the two eclipses are marked with vertical dashed lines.}
\label{fig:kic-residuals-iterationss}
\end{figure}

\citet{Maceroni.etal.2014} started their iterative procedure with cleaning the pulsations identified from the harmonic analysis of the data outside the eclipses, then binary modelling, followed by harmonic analysis of the full data range, and repeating the last two steps until convergence. Each of these steps operates on the residuals of the previous step.
We followed the same procedure, with a few differences. 
(\textit{i}) We started with the eclipsing binary (EB) modelling of folded and binned data {together with the pulsations}, using the parameters determined by \citet{Maceroni.etal.2014} as initial values; 
(\textit{ii}) All pulsations were treated as pure sinusoidal signals, until the binary parameters stopped improving; 
(\textit{iii}) Peaks that were multiples of the orbital frequency were considered as residuals from an incomplete binary modelling, their signals collected and transferred from the pulsation data to the EB data;
(\textit{iv}) After convergence was achieved in 4 iterations, all but the 8 largest frequencies were subtracted as radial modes modulated on the secondary (by being multiplied by the secondary's normalised contribution to the integrated flux, as derived from the binary model), while the 8 largest modes were fitted with an eclipse mapping reconstruction described in Section~\ref{sec:em}. Using this result we performed a last iteration to refine the parameters, specifying the fifth (as the last) iteration.
This latter was a significant improvement, since the bulk of the pulsation signal is contained in these frequencies. It resulted in a dramatic reduction of the residuals in the last iteration, as illustrated in Figure~\ref{fig:kic-residuals-iterationss}.

We omitted the undetermined modes of the rest of pulsations because they do not not impact this analysis.
The only exceptions are probably the hidden modes which have small amplitude when uneclipsed but large amplitude during the eclipses, especially in the rare case of asymmetric modulations. However, we neglected them during the disentangling, because the achieved residuals suggested that only a few such modes may present.

All modes were assumed on the secondary because the diminishing of the residuals during the secondary eclipses indicated so, at least for the largest amplitudes. This was also the conclusion of \citet{Maceroni.etal.2014} (although, as detailed below in Section~\ref{sec:puls-source}, \citet{Manzoori2020MNRAS.498.1871M} argued that the second frequency has to be the fundamental frequency of the primary; we discuss the source of pulsations in that Section). The assumption is certainly not entirely true, as there are non-negligible residuals during the primary eclipses too.

\subsection{Enhancing the binary model}

We used the \texttt{PHOEBE} code \citep{Prsa_phoebe} for fitting the binary model. 
For this purpose the current light curve data were folded within one orbit, and binned with a resolution of 0.0005 between in the phase intervals $[-0.25,-0.21]$ and  $[-0.14,0.14]$ covering the eclipses, and 0.005 elsewhere. The binning also yielded formal point to point errors which were used during the fit.

As initial parameters for the binary model fitting, we used the binary solution determined by
\cite{Maceroni.etal.2014}.
and the adjusted parameters were: the inclination $i$, the surface potentials $\Omega_{1,2}$, the eccentricity $e$, and the argument of periastron $\omega$. All the other values were kept at their values of the initial solution. In particular, we had no reasons trying to re-adjust quantities that had already been determined from spectroscopic data by \cite{Maceroni.etal.2014}. The gravity darkening and albedo coefficients were kept fixed at their theoretical values for radiative envelopes ($g=1$, $A=1$, \citealp{vonZeipel.1924}). The limb darkening coefficients of a square root law corresponding to the \textit{Kepler} transmission function were set to be internally interpolated by \texttt{PHOEBE} for the given atmospheric parameters. 

The derived binary parameters are presented in Table~\ref{tab:binary_pars}, compared with the previous result of \citet{Maceroni.etal.2014} as reference. Our results are quoted with one more decimal digit in order to make it clear that all of them were fitted, and not adopted.

The values differ from the reference values more than the quoted errors in the latter; we think the reason lies in the different handling of the modulated pulsations (radial versus mapped modes). Our primary reason was the removal of the binary signal with the highest possible accuracy -- better than the amplitude of the pulsations to be mapped; otherwise the residuals would be regarded as modulation signal and totally distort the mode mapping results.
Errors in the geometric quantities are better tolerated by the mapping algorithms \citep{Biro::DEM}. 
For this reason we did not attempt to derive uncertainties for the determined parameters. 

\begin{table}
\centering
 \begin{tabular}{lcc}
  \hline
  Parameter & \citet{Maceroni.etal.2014} & This work\\
  \hline
  $i$ & $88.176$            & $88.1942$\\
  $e$ & $0.465$            & $0.46502$\\
  $\omega$ [$^o$] & $21.61$  & $21.4000$\\
  $R_1$ [$R_\odot$] & $3.45$ & $3.465$  \\
  $R_2$ [$R_\odot$] & $3.05$ & $3.000$  \\
  \hline
 \end{tabular}
 \caption{The orbital and geometric parameters of KIC~3858884.}\label{tab:binary_pars}
\end{table}

\subsection{Extraction of pulsation frequencies}
\label{sec:puls-analysis}

In our presented Work, we generally report the frequencies in units of the orbital frequency, $f_\text{orb}=1/P=0.0385\,\text{d}^{-1}$, just as time is measured in units of the orbital period during the eclipse analysis. In the Table~. shown in Appendix, where the first 55 frequencies are listed, a separate column indicates the frequencies in the more conventional c/d unit.

The harmonic analysis of the current pulsation signal was repeated at every iteration as part of the disentangling procedure.
The data were the residuals after the current binary signal was interpolated to the original light curve points and subtracted. The original data consisted of $358\,240$ points, corresponding to a Nyquist frequency limit of $19\,082\,f_\text{orb}$ (or $735\,\text{d}^{-1}$).%
We did an initial analysis on the original data set with full time resolution, in order to check for the highest frequency signal present. It was found that there is no peak in the Fourier spectrum above $580\,f_\text{orb}$ (or $22.3$~d$^{-1}$). Therefore we undersampled the data by 20, i.e. kept every twentieth point, to reduce the unnecessarily high time resolution. This lowered the number of points to $17\,912$ and reduced the Nyquist limit to $750\,f_\text{orb}$ (or $29$~d$^{-1}$). The undersampling does not affect the frequency resolution which is proportional to the inverse of the total time base of the data set (be it the Rayleigh or Kallinger resolution, \citealp{kallinger.2008A&A...481..571K}).

All the eclipses (both primary and secondary) were excluded from the harmonic analysis in this section. Their presence cause more trouble than the periodic gaps due to their masking -- real side peaks around the frequencies of all modulated pulsations versus side peaks in the spectral window (the latter is already affected by the large gap of 15 months between Q2 and Q8 anyway). However when the locations of the individual modes were investigated (Section~\ref{sec:results}), we performed additional harmonic analyses with exactly the same configuration, but one or the other of the eclipses also being involved.

We used \texttt{SigSpec} \citep{SigSpec.2007A&A...467.1353R} to detect and extract the frequencies from the data. It is an automated program which operates on the basis of probabilistic formulation of signal detection, and uses the spectral significance (or negative logarithm of the false alarm probability) to select the peaks. We adhered ourselves to the usual threshold of S/N=4 which corresponds to a significance limit of $\sim 5.47$ \citep{SigSpec.2007A&A...467.1353R}, even though, as noted by \citet{Maceroni.etal.2014}, the real useful limit for the present data is probably as high as 12. Our choice was driven by the desire to obtain a disentangling as perfectly as possible, taking into account every signal, but not belonging to the eclipses.

As a precaution and as a convenience in the observational asteroseismology, we applied a classic filtering mechanism in order to select the real pulsation frequencies. If nonlinear processes are present in the stellar pulsation, an unknown (or rather not well known) mechanism occurs, which result is the so called combination frequencies. This is defined by the following expression: 
\begin{equation}
    f_{ij} = c_i f_i + c_j f_j \pm \Delta f
\end{equation}
where $f_i$ and $f_j$ are the parent frequencies, $c_i$ and $c_j$ are integer numbers and $\Delta f$ is the Rayleigh frequency resolution, the inverse of the time base of the full data set \citep{papics_freq2012}. We wrote a Python script which aim is to sort the frequencies found by SigSpec to the proper group, depending on whether a frequency is equal to a combination frequency with the threshold of Rayleigh-resolution of the data series.

\section{Identification of the source stars of the pulsations}
\label{sec:puls-source}

Needless to say, the performance of mode identification requires the right frequencies modelled on the right stars.
In this regard, there is some controversy among the published results about KIC~3858884.
\cite{Maceroni.etal.2014} concluded that at least the strongest pulsations are most likely originate on the secondary, indicated by the behaviour of the residuals in the Rossiter-McLaughlin effect and in the light curve cleaned from pulsations.
On the other hand, \cite{Manzoori2020MNRAS.498.1871M} argued that the two largest pulsations of similar frequency and amplitude (cf. Table~\ref{tab:freqlist}) are the fundamental modes of the two similar components, and provided support by computing time delays for the two pulsations, using the phase modulation method of \citet{Murphy.etal.2014}, showing oppositely varying phases for the two pulsations (Appendix B. of cited work).
If they are indeed the fundamental modes, it seems logical to assume that they originate from two stars with similar physical properties. 

In order to settle the question, we first investigated these two frequencies and then attempted to determine the source for the other frequencies using 
various methods.

\begin{figure}
\begin{center}
\includegraphics[width=\columnwidth]{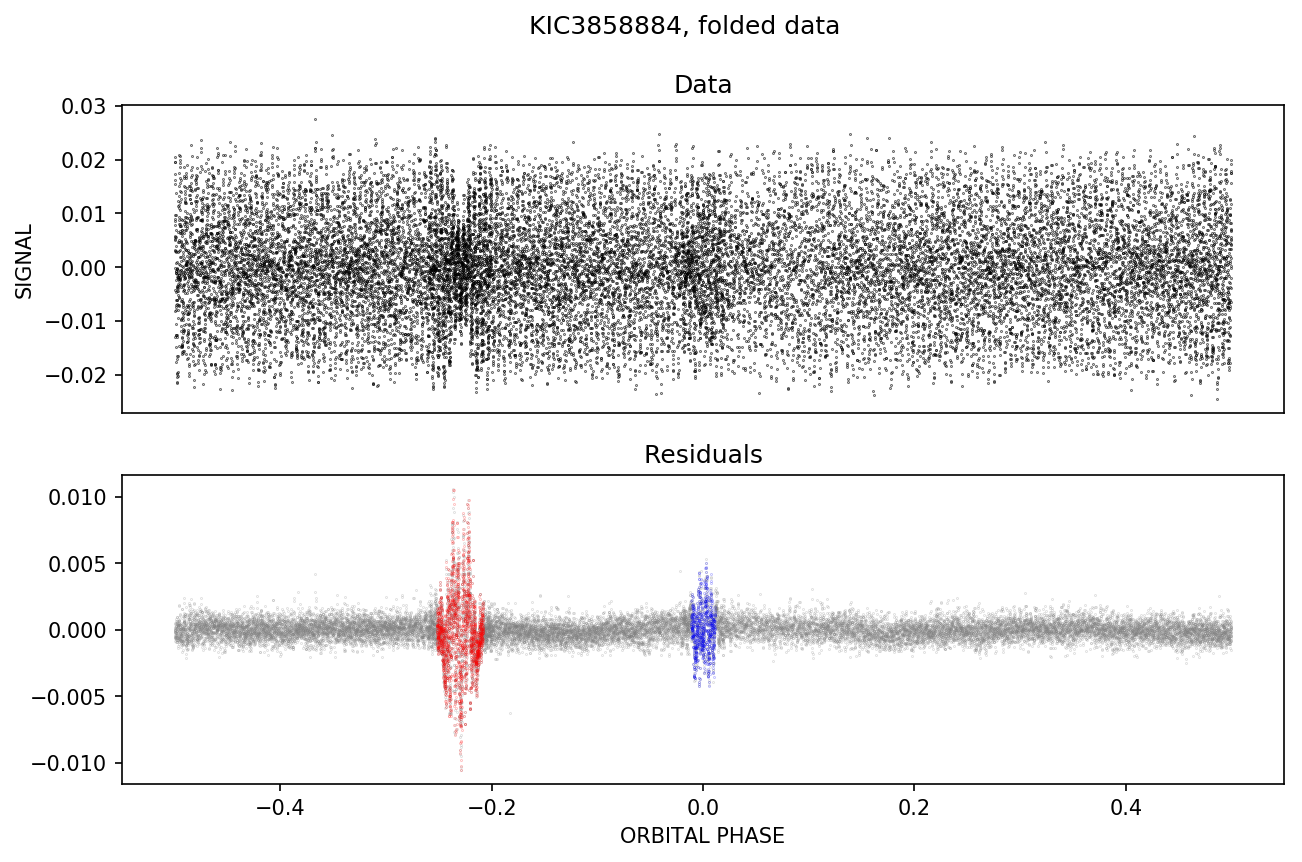}
\end{center}
\caption{Residuals of KIC~3858884 folded with the orbital phase. 
         \textit{Top}: after subtracting the binary model;
         \textit{Bottom}: after also subtracting all the pulsations {modelled as pure, uneclipsed sinusoidal signals}.
         Blue and red colours mark the sections affected by the primary and secondary eclipses, respectively.
         Only every second data point is drawn for a better  visibility.}
\label{fig:kic-residuals}
\end{figure}

Figure~\ref{fig:kic-residuals} shows the residuals of the light curve at various stages,
folded with the orbital period. The top panel displays the pulsation signals after the subtraction of the EB model. The diminishing of the pulsation amplitudes during the secondary eclipses is evident. There is also some modulation going on during the primary eclipse phases, but is less pronounced. The bottom panel displays the residuals after also subtracting all the pulsations, but as pure harmonic signals, without any modulation anywhere. Therefore, the more {real} modulation is going during an eclipse, the larger its residuals are.
It is clear that if F1 and F2 were fundamental radial modes of two different stars, then the observed pulsations would experience modulations of comparable magnitude at the two eclipses. Yet, we see a very pronounced effect during the secondary eclipses and a very small one during the primary one.

\begin{figure}
\begin{center}
\includegraphics[width=\columnwidth]{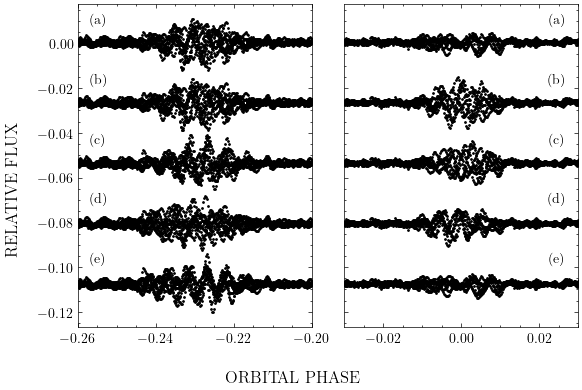}
\end{center}
\caption{Residual investigation for identifying the sources of F1 and F2. All the other frequencies were subtracted as unperturbed sinusoidals. From top to bottom: 
(a) -- both F1 and F2 are also sinusoidals; 
(b) -- both are on the primary; 
(c) -- F1 on primary, F2 on secondary;
(d) -- reverse of (c);
(e) -- both are on the secondary.
}
\label{fig:puls-source-residual}
\end{figure}

We experimented exclusively with the locations of F1 and F2. For this, we got rid of all the other pulsations as unperturbed harmonic signals. Then we subtracted also F1 and F2 under all possible variants of their location, and compared the residuals. Figure~\ref{fig:puls-source-residual} shows the results. Whenever either frequency is assumed on the primary, the residuals increase dramatically. The lowest residuals are attained when both are modulated on the secondary (bottom plot in the Figure).

The above findings still do not exclude that one of the two pulsations is on the primary: it could be different from a radial mode, thus causing more subtle modulations, as illustrated in Figure~\ref{fig:synthetic-modulations}.

\begin{figure}
\begin{center}
\includegraphics[width=\columnwidth]{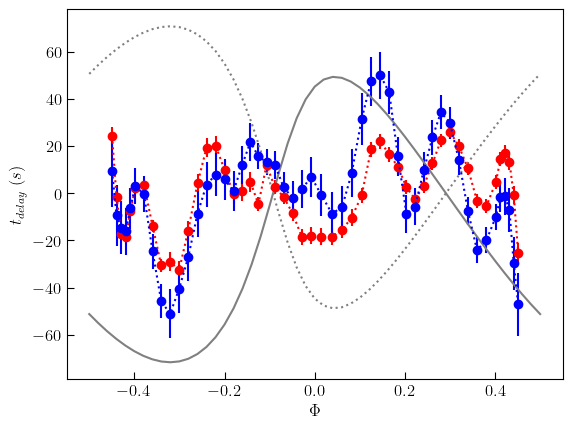}
\caption{Time delays of F1 and F2 computed with the modified phase modulation method (see text for the details). The blue and red filled circles correspond to F1 (f~=~187.65~$f_\text{orb}$) and F2 (f~=~193.95~$f_\text{orb}$), respectively. The continuous and dotted grey lines show the theoretical time delays computed by the second generation \texttt{PHOEBE-2} .}
\label{fig::pm-investigation}
\end{center}
\end{figure}

\citet{Manzoori2020MNRAS.498.1871M} used the phase modulation method to prove that F1 and F2 show phase variations of opposite sign. In order to clear up the issue, we have also performed a PM analysis. The classic method, which works in the time domain (without the knowledge of the orbital period) found periodic phase variations for F1 and F2 that were running together, but with a period different from the known orbital period.
Therefore we changed to the orbital phase domain with the known period, and sliced the times series into smaller segments according to the orbital phase rather than time.
The rationale was that there may be other stellar or non-stellar processes influencing the fit of initial phases, however our main aim was only to determine the sources of the frequencies. 
The eclipse phases were omitted from the analysis in order to avoid the complications related to modulated pulsations.
We then fitted the first eight frequencies constraining their frequencies and out of eclipse amplitudes, adjusting the phases only.
The modified method met our expectations, we could obtain firm evidence that F1 and F2 are really on the secondary component (see Fig~\ref{fig::pm-investigation}). For these frequencies, the obtained time delays are in fair agreement with the theoretical values, which were computed from line-of-sight distances modelled by second generation of Phoebe \citep{Prsa2016ApJS..227...29P}. 
As for the other frequencies, the results were inconclusive, probably due to their lower amplitudes.

Finally, we performed a so-called {double} eclipse mapping of the first 8 frequencies, where mode pulsations on both stars are mapped simultaneously, using the data from both eclipses. 
If a pulsation is assumed on a star, but has no associated modulations in the data set during the eclipses of that star, then its surface pattern is expected to receive a negligible power.
Therefore, we assigned the same 8 frequencies on both on the primary and secondary, and let the mapping procedure decide about their proportions on each star, in terms of amplitude. Because the emphasis was the separation of the modulations rather than providing sensible pulsation maps, no symmetry was assumed at all about the solutions (also because at this stage we have no information on the orientation of the pulsation axes); instead, a {most uniform} solution was sought.
The amplitudes of each reconstructed pattern, computed outside the eclipses, are shown in Table~\ref{tab:kic-double-em}, and indicate that the majority of the most dominant modes, including $f_1$ and $f_2$, come from star 2, although the disambiguation becomes less reliable when going towards the weaker modes. 
In fact, F9 and F10 may as well reside on the primary. We note in passing that F9 is the almost exact double of F10, so the two pulsations may indeed have a common origin, suggesting the presence of some nonlinear modes as well.

\begin{table}
\centering
\begin{tabular}{clllr}
\hline
Num. & Freq [$f_\text{orb}$] & Star & Amplitude & Phase [$\deg$] \\
\hline\hline
F1 & 187.65 & 1 & 0.0012           & -155.1 \\
   &             & \textbf{2} & \textbf{0.0086}  & -150.4 \\
F2 & 193.95 & 1 & 0.00122          & 53.5 \\
   &             & \textbf{2} & \textbf{0.00802} & 89.0 \\
F3 & 255.30 & 1 & 0.00027          & -94.5 \\
   &             & \textbf{2} & \textbf{0.00183} & -4.1  \\
F5 & 174.81 & 1 & 0.00033          & -110.7 \\
   &             & \textbf{2} & \textbf{0.00132} & -162.6 \\
F6 & 247.04 & 1 & 0.00052          & -63.2 \\
   &             & \textbf{2} & \textbf{0.00150} & 166.7 \\
F8 & 304.31 & 1 & 0.00026          & 25.2 \\
   &             & \textbf{2} & \textbf{0.00081} & 53.2 \\
F9 & 382.15 & 1 & 0.00038          & -107.8 \\
   &             & \textbf{2} & \textbf{0.00058} & 0.3 \\
F10 & 191.08 & \textbf{1} & \textbf{0.00362} & 102.5 \\
   &             & 2 & 0.00326          & -84.8 \\
\hline
\end{tabular}
\caption{Result of the double eclipse mapping for KIC~3858884 in terms of the integrated out-of-eclipse signal properties attributed to each component. Bold letters mark the larger amplitude for each frequency.}
\label{tab:kic-double-em}
\end{table}

\begin{table}
    \centering
    \begin{tabular}{cc|ccc}
       ID   & Frequency ($f_\text{orb}$)  & PM & DEM & Note\\
       \hline
       \hline
       F01 & 187.65 & SEC & SEC &\\
       F02 & 193.95 & SEC & SEC &\\
       F03 & 255.30 & o & SEC &\\
       F04 & 194.96 & -- & -- & F01 + $f_\text{orb}$\\
       F05 & 174.81 & o & SEC &\\
       F06 & 247.04 & o & SEC &\\
       F07 & 381.60 & -- & -- & F01 + F02\\
       F08 & 304.31 & o & SEC &\\
       F09 & 382.15 & o & SEC &\\
       F10 & 191.08 & o & PRI &\\
       \hline
    \end{tabular}
    \caption{Summary of the pulsation source analysis for the KIC~3858884 system. PM and DEM stand for Phase Modulation and Double Eclipse Mapping methods, respectively.}
    \label{tab:double_em_freq}
\end{table}

In our opinion, the results (see Table~\ref{tab:double_em_freq}) of the investigations detailed above provide a firm evidence that F1 and F2, as well as most of the significant pulsations, are located on the secondary star. There is also some evidence that the five of other dominant frequencies are identified to be from the secondary, and the one with the lowest amplitude (F10) among them from the primary star.

\section{Mode identification}
\label{sec:modeid}

\subsection{Methods}

In the linear adiabatic approximation the surface pulsation patterns of the oscillation eigenmodes of a spherically symmetric star can be described in terms of surface spherical harmonic functions $Y_\ell^m\left(\theta, \varphi\right)$ in the coordinate system tied to the symmetry axis of the star \citep{Cox1980}. The two numbers  $\ell$ (degree) and $m$ (azimuthal order) are the two surface mode numbers identifiable with various photometric or spectroscopic techniques (usually only $\ell$ can be determined).
Rotation and tidal interaction cause distortions in the amplitude distribution of these eigenmodes in addition to frequency splittings (\citealp[][]{Asteros_book}), leading to equatorial concentration \citep{Reese2009} or mode trapping \citep{Handleretal_2020}, among others.
Therefore the use of spherical harmonics may not always be appropriate or justifiable. Nevertheless, for the study of intrinsic stellar oscillations in binaries where these perturbing effects are small, they can still be a useful approximation. In accordance, we have employed two approaches in modelling the surface patterns. The first, Dynamic Eclipse Mapping method (hereafter DEM), reconstructs general surface pulsation patterns with the possibility to impose certain symmetry properties on them, from which the mode numbers can be inferred either by topological resemblance or by quantitative analysis of the amplitude and phase profiles. The second, Direct Fitting (hereafter DF), works with the usual surface spherical harmonics as basis functions, and looks for the best matching set of mode numbers either by sequential probing of all possible cases, or by a stochastic sampling of their parameter space. It also has the potential to detect a possibly tilted pulsation axis, which cannot be excluded in such a wide binary system.

Both methods rely on the assumption that the resulting time-dependent surface pattern can be approximated by the linear superposition of the individual oscillating patterns. 
For small amplitude pulsations this is the usual assumption made by all other analysis approaches too, on the basis that small changes in the local temperature can be added up linearly, and also induce a linear change in the local brightness around its value determined by the equilibrium value of the local effective temperature at first approximation.
Also, they were constructed to only study intrinsic, free pulsations in binaries. The analysis of tidally locked pulsations with rotating symmetry axes, or trapped pulsations with unevenly surface distribution of the amplitudes, is beyond their scope. Although future extensions could make these cases partially tractable, that is, in turn, beyond the scope of our paper.

\subsubsection{Eclipse Mapping}

Eclipse Mapping (EM) is basically an image reconstruction technique originally conceived for mapping the brightness distributions of accretion discs in eclipsing cataclysmic binaries \citep{em_original.horne.1985} an later in more regular systems \citep{em_eb.collcameron.1997}.
Eclipse events are effectively sampling the surface of the star undergoing the eclipse, convolving its surface brightness distribution into the flux variations over time, i.e. the light curve. Eclipse mapping methods employ regularised inversion techniques to restore the brightness distribution from the light curve.
(We are aware that there is some divergence in using the term "eclipse mapping" across the scientific community; some communications use it to name the phenomenon itself, others refer to the reconstruction method. We follow the tradition of the first papers on the topic, and use the term to name the analysis method.)

\citet{Biro::DEM} constructed a variant of EM extended to reconstruct periodic time-dependent surface patterns, primarily pulsations. For every involved frequency it furnishes two maps (the so-called 'cosine' map $A\cdot\cos\phi$ and its 'sine' counterpart, $A\cdot\sin\phi$), leading to amplitude and phase maps $A$ and $\phi$. The convolution causes a 
significant loss of information on the patterns themselves; however, with reasonable assumptions about their nature, in terms of symmetries, they can be restored to an extent that allows mode identification through the analysis of their the nodal line structure.
The assumption is that the amplitude depends on the co-latitude only; similarly the phase depends on the longitude only, and {linearly}. This approach also allows for  deviations from spherical harmonic caused by rotational distortion.
Since its original implementation, the DEM algorithm has received a significant improvement: the use of a hidden image space, connected to the real image space via a fuzzy pixel-based intrinsic correlation function. This separation of regularisation and correlations led to a dramatic improvement in the smoothness and reliability of the reconstructed patterns, which previously rendered the method awkward for real applications.

Both the application of symmetry requirements and the inference on the mode numbers require the knowledge of the symmetry axis, fixed in space. In close or moderately wide systems it is customary to assume an orbitally aligned rotation axis, to which the pulsations are aligned, unless there is
evidence indicating otherwise (e.g from the Rossiter-McLaughlin effect, \citealp{diher.rme.2009}). In wide systems like KIC~3858884 there is less guarantee that this holds. 
Fortunately the Direct Fitting method has the potential to estimate an orientation from the pulsation data.

\subsubsection{Direct Fitting}

In addition to Eclipse Mapping, \cite{Biro2013EAS....64..331B} also reported on another algorithm which models the surface patterns explicitly as spherical harmonics $Y_\ell^m\left(\theta, \varphi\right)$ defined in a coordinate system tied to the symmetry axis of the pulsation.
For a mode with known $\ell,m$, the only free parameters are its amplitude scale and (initial) phase. If its out of eclipse amplitude and phase in the integrated flux as determined from time series analysis are applied as additional constraints, even these are uniquely determined, and the problem reduces from a fit to a simple evaluation of the match between synthetic light curves and data, for all reasonable modes $(\ell,m)$. The idea is to test all possible partitionings of the modes, up to a certain maximum $\ell$, and select the combination giving the best fit to the data (Direct Fitting of $Y_\ell^m$, DF).
However, as the number of simultaneous modes increases, the number of combinations to be tested also increases as $\left(\ell+1\right)^{2 N}$, where $N$ is the number of frequencies and $\ell$ stands for the maximum degree, which makes the process computationally challenging. For example, for 8 frequencies and degrees up to $\ell=3$ it would mean more than 4 billion cases to investigate (the limit of selectivity is on $\ell-|m|$, rather than $\ell$ itself).

There is an existing semi-solution, which is a variant of DF. The algorithm is similar to a method in time series analysis known as frequency whitening, just performed with $Y_\ell^m$-s, calling hence DFCLEAN in the following. The idea is to search for the best fit per frequencies instead of simultaneously on all frequencies, while the unknown ones are kept as radial and the identified ones with determined ($\ell$,$m$) mode numbers. Although the search is way faster in that way than the DF, this algorithm probe much lower number of possible nonradial mode configurations.

To overcome this issue, we developed a program which performs a Markov Chain Monte Carlo (MCMC) search using the direct fit (evaluation) as its core program \citep[][YLMCMC]{Bokon.mcmc.2020BlgAJ..33...47B}. 
A simple Metropolis-Hastings algorithm is used for sampling. The modes are treated as {categorical variables}, discrete variables with special ordering.
After experimentation, the best ordering of the modes proved to be according to the deviation of their modulation curve -- in the context of the investigated binary model --from that of the radial mode (0,0), in terms of the sum of square of the residuals ($\chi^2$). 

In our runs the mode degree was limited to $\ell \le 3$. Modes with larger $\ell$ have cancellations so large that they are invisible outside the eclipses, therefore undetected.
The starting point of the Markov chain was in every run the all-radial modes case, and all modes are assigned uniform prior probabilities.  
The resulting joint posterior probability distribution of $\ell$ and $m$ parameters for all modes is then used to draw conclusions.

The modelling with spherical harmonics also opens up the possibility of inferring a most plausible value for the pulsation axis, being the polar axis of the coordinate system in which the pulsation patterns can be described as spherical harmonics. As with EM, DF also requires the knowledge of this axis when fitting individual modes. However, when a spherical harmonic is viewed from another coordinate system with a different polar axis, it can be described as the linear combination of all spherical harmonics of the same degree $\ell$, defined in the new system:

\begin{equation*}
    Y_{\ell}^{m}(\theta,\phi) = \sum_{m'=-\ell}^{\ell} {\cal D}^{\ell}_{m' m}(\alpha,\beta,\gamma) \cdot Y_{\ell}^{m'}(\theta,\phi),
\end{equation*}

where ${\cal D}^{\ell}_{m' m}(\alpha,\beta,\gamma) = e^{-im'\alpha}\,d^{\ell}_{m'm}(\beta)\,e^{-im\gamma}$ are the elements of the Wigner rotation matrix which is a function of the Euler rotation angles $\alpha$, $\beta$ and $\gamma$ of the transformation from the current system to that of the pulsation axis. $\alpha$ is the azimuth of the axis, clockwise from the meridian facing the observer and around the orbital axis; $\beta$ is its tilt from the orbital axis; the phase shift $m\gamma$ of the third angle is incorporated as a common phase shift of all the $Y_{\ell}^{m'}$ components, and does not appear explicitly in the analysis.
Therefore it is possible in principle to infer a best estimate of the pulsation axis in two steps. First a set of $\ell$-multiplets 
${\cal Y}_{\ell}(\theta,\phi) = \sum_{m'} c_{\ell m'} Y_{\ell}^{m'}(\theta,\phi)$ is assumed for each pulsation, and the DF procedure now partitions various $\ell$-s (or samples their parameter space). Then the Wigner coefficients of all modes are solved for a common value of $\alpha$, $\beta$ via a numerical optimisation.
Tests performed in \citet{Biro::DF} were encouraging, showing that the orientation of the pulsation axis may be recoverable even in the presence of rotationally distorted modes.

The road map of analysis is then as follows. First a Direct Fitting of multiplets is performed to estimate the obliquity of the pulsation axis. The derived orientation of the axis is then used, first to do an Eclipse Mapping which furnishes reconstructed surface patterns, and allows estimates for mode numbers; and secondly, another Direct Fitting of single modes to infer mode numbers in the frame of the spherical harmonics model. The comparison of the two sets of mode numbers may shed light on deviations of the pulsation patterns from spherical harmonics, or point to limits of capability of the method for the particular binary system.

There is a limit on the number of simultaneously mappable modes, as revealed by tests performed previously (\citealp{Biro::DEM}). Single modes can be successfully reconstructed even from a single sampled eclipse. Multiple modes require multiple eclipses in order to be properly separated during the mapping. As a rule of thumb, it was found that, for reasonably precise data like those from space telescopes, the number of simultaneously restorable modes will not be larger, possibly even less, than the number of the eclipses involved. Larger noise in the data can be compensated by collecting observations of more eclipses.

The algorithms were also found to be less reliable when more than a few modes were present {and} some pulsations had much smaller amplitudes than others. In this case, the smaller amplitudes tend to be improperly reconstructed. To deal with this shortcoming, we analysed the pulsations in groups or packets of roughly similar orders of amplitude, in decreasing order.
All the modes to be mapped were temporarily cleaned as radial modes with uniform amplitude and phase patterns, affected only by decreasing variation of their amplitudes proportional to the fractional contribution of their host star to the integrated flux; in our study, all the studied modes were located on the secondary component. All the modes already mapped were properly subtracted with their fitted modulations.

\subsection{Results}
\label{sec:results}

Once the origins of the frequencies were clarified, the actual mode-identification task could begin. 

For the mapping procedures the data were resampled to uniformly spaced points with more amenable resolution in time. They covered the secondary eclipses and some of their surrounding out of eclipse portions. The same set points in terms of fractional (or photometric) phase was used for all the individual eclipses. This speeds up the mapping procedures: an eclipse kernel covering a single secondary eclipse can be replicated for all the eclipses, shifting its temporal factors properly in time. Quadratic interpolation was sufficient for the resampling, given the density of the original data points. Each section was covered by 801 points from -0.28 to -0.18 in orbital phase. This covers the secondary eclipses between -0.25 and -0.23 as well as two additional sections on the two sides; the latter are included to constrain the uneclipsed amplitude and phase characteristics of the modes to be fitted. The 11 secondary eclipses contained in the data are thus covered by 8811 individual data points. (Incomplete eclipse phases received masked values for their missing time points, which are ignored by the algorithms.) 

We first attempted this for the frequencies listed in Table~\ref{tab:double_em_freq}, omitting the combination frequencies. 
A DF in multiplet mode yielded angles $\alpha\sim 32^\circ$ and $\beta\sim 30^\circ$ for the symmetry axis.
However, both the EM and DF runs were essentially unsuccessful, regardless of whether all the frequencies were handled together, or grouped in 4+4 according to similar amplitudes, or cleaned one by one (using DF for the latter). The achieved fits to the data were only marginally better than the all-radial case. In the case of the EM, tightening the fit (in terms of the sum of squared residuals, $\chi^2$) only increased the noise in the images but did not reduce the residuals under the eclipses, signalling that the algorithm is fitting noise rather than structure.
For reference, the runs were repeated with perfectly aligned pulsation axis, with the same result.

\begin{figure*}
\begin{center}
\includegraphics[width=1.95\columnwidth]{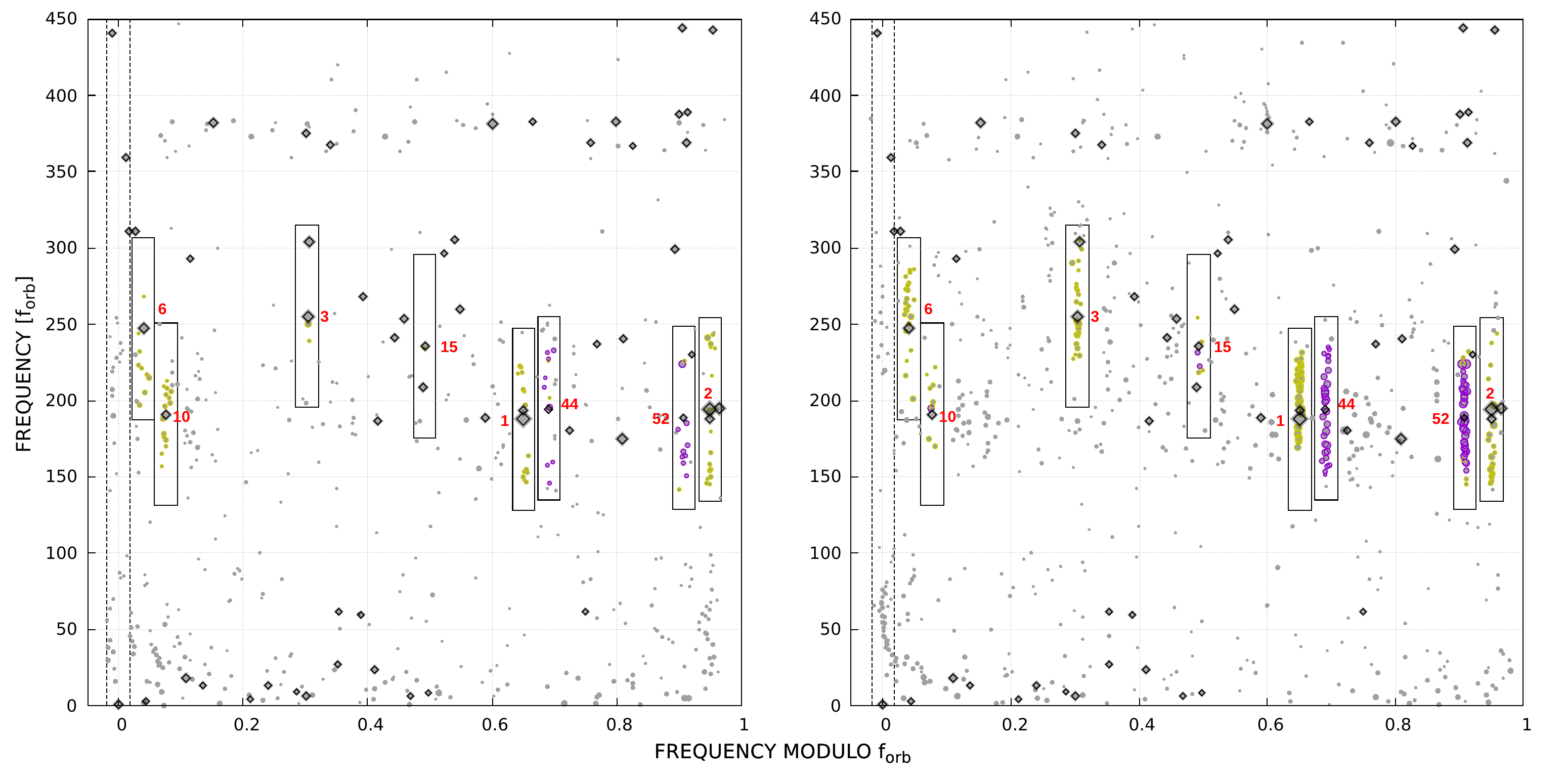}
\caption{\'{E}chelle diagrams of frequencies for the data including  primary eclipses (\textit{left} panel),  secondary eclipses (\textit{right} panels). Grey filled circles mark the individual frequency peaks intrinsic to the actual dataset. The common peaks (present in all datasets, including the one without any eclipses) are shown with the bold diagonal squares. The numbered peaks are those for which a significant amount of side peaks were found in either of the cases. The side peaks are colour coded according to the ratio of their amplitudes to that of their central peak: ratios lower than 5\% are shown in yellow, the larger ones in violet. (The threshold only serves visualisation purposes). All symbol sizes are proportional to their amplitudes on a logarithmic scale. The boxes show the regions in which side peaks were searched for; their widths is equal to the Rayleigh frequency resolution ($\sim 0,036$), and extend vertically by $50\,f_\text{orb}$ in both directions. Only F10 was found to have significant modulation during the primary eclipses. The dotted box centered on 0 shows the location where orbitally resonant frequencies and remnants of an incompletely subtracted binary signal are expected.}
\label{fig::kic-echelle}
\end{center}
\end{figure*}

This called for a more thorough investigation regarding which modes are really involved in the modulations.
The nature of the residuals pointed to the possible presence of unaccounted {hidden} modes contaminating the data. These are modes with surface patterns that yield nearly perfect mutual cancellation outside the eclipses, but get amplified during the eclipses due to the symmetry breaking. (1,0), (2,1) or (3,1) are such modes when viewed nearly edge-on, like in the present system.

Fortunately the amount and quality of the short cadence data for KIC~3858884 allowed us to the search for modes affected by the eclipses in the frequency spectra.
Eclipses modulate the amplitudes of the individual pulsation modes with the periodicity of the orbital motion. It results in the development of side peaks in the frequency spectrum around the frequency of the affected mode, with spacing equal to the orbital frequency, $f_\text{orb}$. Tidal perturbations also induce frequency splitting with the same spacing, although there were cases where the excited pulsations were close to the harmonics of orbital frequency \citep{Bowman2019ApJ...883L..26B} instead of the splitting of actual frequencies.
(In addition, unaccounted base intensity variations may also introduce such side peaks in a sophisticated manner: residual base intensity variations themselves cause peaks at exact multiples of the orbital frequency, but the associated variations in all pulsation amplitudes of the star in question implied by variations in the projected area of the stellar disc show up as the same side peaks around the main peaks.)
However, these modulations are always present, while the eclipse modulations only occur if the eclipse phases in question are involved in the analysis.
Following this logic, frequency identification can be performed from a masking procedure by excluding eclipses completely or by excluding one or more overlaps.  For this approach, we made of use the SigSpec \citep{SigSpec.2007A&A...467.1353R} program, which has automatic prewhitening process searching the most probable peaks in the Fourier-spectra instead of the highest amplitude, like Period04 does. Thus, in addition to the three Fourier spectra, we obtained three determined frequency-sets, on the latter of which we can perform cross-identification.

Following this logic, two additional \texttt{SigSpec} runs were made, using the same setup as before, but on datasets extended to include either the primary or the secondary eclipses, respectively. We refer to the original spectrum presented in Section~\ref{sec:puls-analysis} 'noe', and these two spectra as 'pri' and 'sec'.
(The three datasets have the out of eclipse points in common.) 
The significance limit was lowered for the latter two to 3, in order to detect as many side peaks as possible.
The peaks were then examined on \'echelle diagrams with the folding frequency of $f_\text{orb}$, which is 1 in our case (all frequencies are reported in units of the orbital frequency). Fig.~\ref{fig::kic-echelle} shows the two diagrams for the 'pri' and 'sec' cases, together with the peaks common to all three spectra. 
We also cross-identified the peaks across the three cases. In doing so, we took the Rayleigh frequency $f_\text{R} \approx 0.036$ as a reference value for the precision of the peak frequencies, although the matching of the peaks turned out to be much tighter, within about 1/4 of the quoted precision, in agreement with the heuristic results of \citet{kallinger.2008A&A...481..571K} that the precision also depends on the significance of the peak.
(There were also about 70 peaks found exclusively in 'noe', i.e. they were missing from both 'pri' and 'noe'. All of them have amplitudes lower than $5 \times 10^{-5}$; we suspect they are due to the imperfections imposed by the data quality.)

\begin{figure*}
\begin{center}
\includegraphics{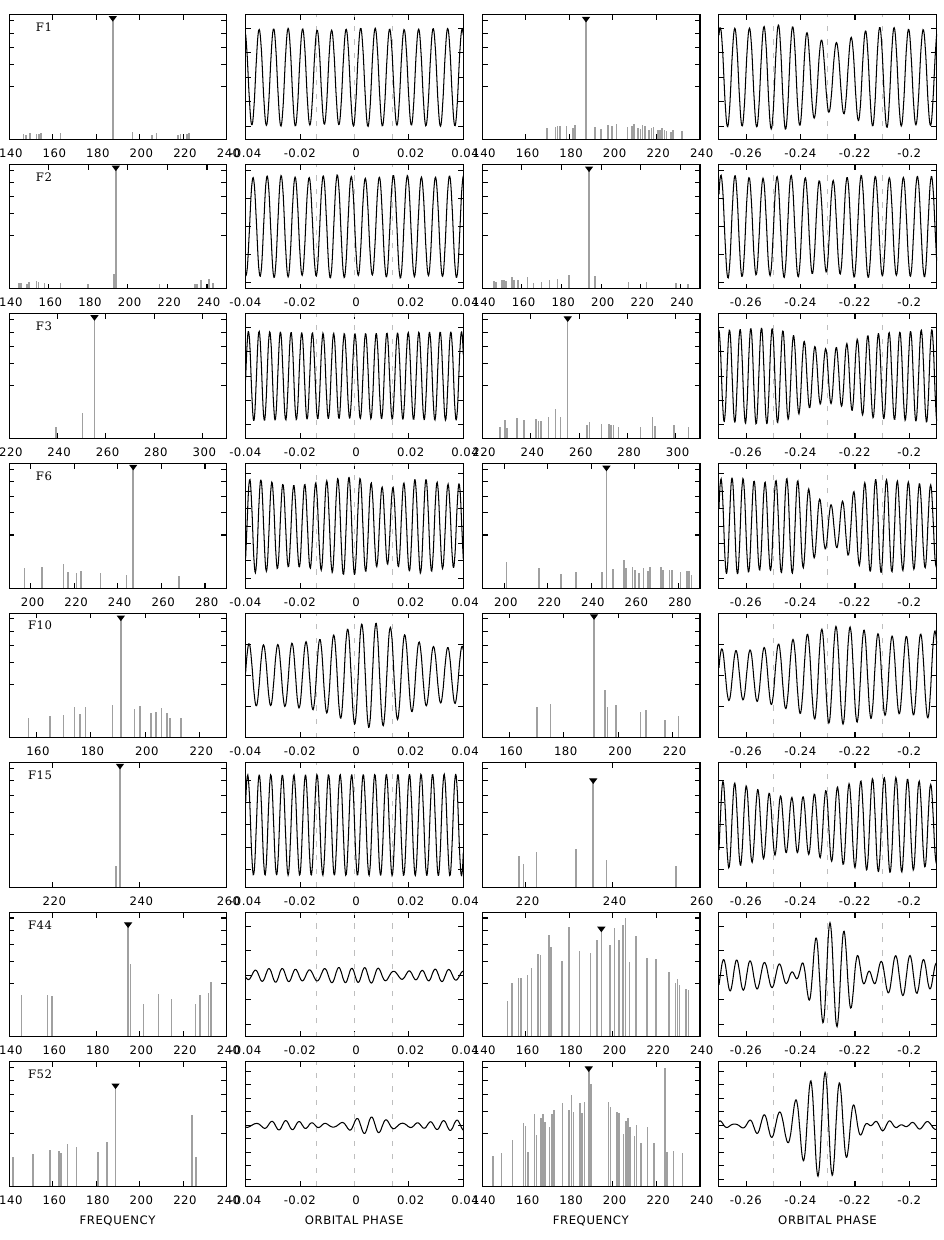}
\caption{The modulations of selected frequencies, as determined from their sidelobes identified in the 'pri' and 'sec' spectra involving primary or secondary eclipses, respectively. Every row contains data for one frequency, marked in the leftmost panels. From left to right: \textit{columns 1 and 2} -- data (peaks + their synthetic signals) for primary eclipses; \textit{columns 3 and 4} -- same for secondary eclipses. Note that the peak amplitudes in the frequency spectra are plotted on a nonlinear -- square root -- scale in order to make the smallest peaks visible. Black upside down triangles mark the central peaks. Dashed vertical lines in the light curve panels (columns 2 and 4) mark the eclipse regions.}
\label{fig::echelle-modulations}
\end{center}
\end{figure*}

It is intriguing to see the very low level of modulations during the primary eclipses. Only a few peaks show notable 'pri' side lobes around them: F10 and perhaps F13 and F4, but the latter two are blended with F1 and F2. The secondary eclipses, as expected, affect a larger number of modes and more strongly, namely F1 and F2 as expected, then F3, F6 and F15, and, unexpected from previous evidence, F44 and F52. 
Figure~\ref{fig::echelle-modulations} shows the modulations of these modes computed by adding up their main peaks and side lobes. F44 and F52 clearly have the characteristics of hidden modes, their amplitudes increasing dramatically during the secondary eclipses, reaching levels equal to the residuals from the first unsuccessful mappings.

Thus from the \'echelle diagrams we updated the set of frequencies to be mapped on the secondary: frequencies 1, 2, 3, 6, 15, 44 and 52. Their characteristics are replicated in Table~\ref{tab:secondary-freqs-2} from the main frequency table in Appendix, for easier reference.

\begin{table}
    \centering
    \begin{tabular}{ccc}
       \hline
       \hline
       ID   & Frequency ($f_\text{orb}$)  & Amplitude \\
       \hline
       F01 & 187.65 & 0.00982  \\
       F02 & 193.95 & 0.00905  \\
       F03 & 255.30 & 0.00186  \\
       F06 & 247.04 & 0.00124  \\
       F15 & 381.60 & 0.00033  \\
       F44 & 304.31 & 0.00014  \\
       F52 & 382.15 & 0.00018  \\
       \hline
    \end{tabular}
    \caption{The final set of frequencies selected for mode identification. All are assumed on the secondary component.}
    \label{tab:secondary-freqs-2}
\end{table}

We generated artificial modulated pulsations for all spherical harmonic modes with $\ell$ up to and including 3, using the current binary parameters and the same time points as used for the time series analyses above. The first frequency F1=$187.65\,f_\text{orb}$ was used for all modes. No noise was added. We then performed \texttt{SigSpec} runs on these {noiseless} data to locate the first 100 peaks in the frequency range 100 to 300 (also in units of $f_\text{orb}$). The simulated data and their spectra are shown in Figures~\ref{fig:synthetic-modulations} and \ref{fig:synthetic-spectra}. The structure of side peaks is clearly visible for all cases. They are shown for positive $m$-s only; for negative $m$ they are mirrored around the central peak. (They are shown on the panels of modulated light curves, though.) We experimented with the number of peaks (going from larger to smaller) required to reproduce the modulated signals to an acceptable accuracy. We found that at least 80 peaks were required for a good approximation. 

For comparison, we located the side peaks associated to the modulated frequencies based on the 'sec' \'echelle diagram. In this procedure the matching tolerance was lowered to $1/4\,f_\text{Ray}$, in order to exclude close but unrelated peaks. Also, only peaks present exclusively in the 'sec' spectrum (i.e. neither in 'noe' nor in 'pri') were considered as candidates (to avoid any spurious side peaks or of possible tidal origin).
The side peaks and their resulting modulated pulsations are shown in Figure~\ref{fig::echelle-modulations} in a similar manner as for the simulated data.

The number of detected side peaks from the real data with noise is much lower than from the simulations. Therefore a meaningful isolation of the individual modulations based on their detected side peaks is beyond possible, even for these extraordinarily precise data. 
Instead, one can go the other way around: subtract the signals of all the peaks of the 'sec' spectrum from the data, {except} the target peaks and their side lobes. We note that the peaks in 'sec' already describe the modulations during the secondary eclipses, therefore they should be interpreted as unmodulated harmonic signals.

\subsubsection{Direct Fitting}
\label{sec:df}

First, we performed a direct fitting of $\ell$-multiplets in order to estimate the orientation of the symmetry axis. The first five modes were used for this purpose. (A seven frequency case has also been done, with close results; however, its following validation run with individual modes or {singlets} $(\ell,m)$, encompassing almost 300 million individual cases, would have taken too much computational times, therefore we gave up on it.)
The best solution, with $\chi^2=1.48$, yielded the $\ell$-numbers (1, 1, 2, 2, 2) for the five frequencies. The fitting of Wigner coefficients on the best parameters gave $\alpha=21.9^\circ$ and $\beta=21.6^\circ$ for the orientation of the rotation axis. This is very close to an aligned configuration. The associated mode numbers turned out to be 
(1, 0), (1, 0), (2,-1), (2, 0) and (2, 1). This came as a surprise, given that modes (1,0) are hidden modes.

To check the validity of these mode numbers, we performed runs of DF, DFCLEAN and YLMCMC, all in singlet mode. Table~\ref{tab:wigfit-df-mcmc} summarises the results. Generally speaking, there is a good agreement in the mode numbers found by these runs with fixed axis, but they differ significantly from those given by the free axis algorithm. 
We have two conjectures regarding the cause of this disagreement.
It could happen if the pulsation patterns can not be appropriately described by spherical harmonics -- although we think it unlikely for nearly spherical stars on such a wide orbit (period almost 26 days), even if it is eccentric.
Or the remnants of the many other unfitted modes, taken into account as uniform patterns affected by the eclipses, are disturbing the data sufficiently so that sensible information on the orientation of the axis cannot be inferred.

All the analyses have been repeated for an aligned axis case ($\alpha=0^\circ$, $\beta=0^\circ$). The results are shown in Table~\ref{tab:aligned-reconstructions}.

\subsubsection{Dynamic Eclipse Mapping}
\label{sec:em}

As with DF, we proceeded with the eclipse mapping analysis in two consecutive steps, with the two sets of frequencies grouped according to similar amplitudes.

In EM the goodness of fit to the data is user-specified in terms of the chi-squared value.
If only white noise is present and its level is correctly estimated, $\chi^2=1$, or a slightly larger value is appropriate. For white noise of unknown level a noise scaling is possible by tuning $\chi^2_\text{aim}$ so that the value of a related, so-called {R-statistic}, becomes close to zero. (The R-statistic, proposed by \citealp{R-statistics.baptista.steiner.1993}, is a measure of correlation between residuals.)
Unfortunately this approach rarely works in practice due to the presence of additional contaminating factors, like the remnants of modulations of the many smaller amplitude pulsations. Some experimenting is required regarding the smallest value of attainable chi-squared without significant distortions in the maps. The first version of EM as described in \citet{Biro::DEM} was very sensitive to noise, which easily propagated into the maps as unrealistic, noisy patterns as the level of fit was being tightened. The introduction of the hidden space solved this shortcoming, allowing only smooth maps, and because of that, they are more immune to such extra noise. There will be a lower limit of chi-squared below which the algorithm cannot go, giving at that point a smooth, but unfeasible map, in the sense that they do not show the typical "rosette" symmetry of the pulsation patterns connected to a symmetry axis. (The smoothness is on a scale related to the spatial resolution of the hidden space, which is covered at an acceptable level by about 55 uniformly distributed vertex points; therefore smooth maps may still be unfeasible.)
Also, given that many different modes are bound to cause nearly the same modulation, as illustrated in Figure~\ref{fig:synthetic-modulations} for the case of the secondary eclipses of KIC~3858884, as our aimed chi-squared was being pushed down, the solutions kept switching between the various nearly equivalent modes (e.g., (0,0),(1,$\pm$1),(2,$\pm$2) are such a group), separately for each mapped mode. Selecting an appropriate chi-squared then unavoidably injects a certain amount of subjectivity into the solution. More precise data and, more importantly, less correlated noise would certainly help in resolving this ambiguity. In the present case we tried to mitigate the subjective factor by first going for the smallest attainable chi-squared value, and then relaxing it until the maps become feasibly. 
It was comforting to see that at least the \textit{signs} of $m$ remained consistent for each map throughout the iterations.

We note that the values of $chi^2$ attained by the DF fits can only serve as upper limits for EM, because they assume the more restrictive spherical harmonics.

\begin{figure*}
    \centering
    \includegraphics[width=\textwidth]{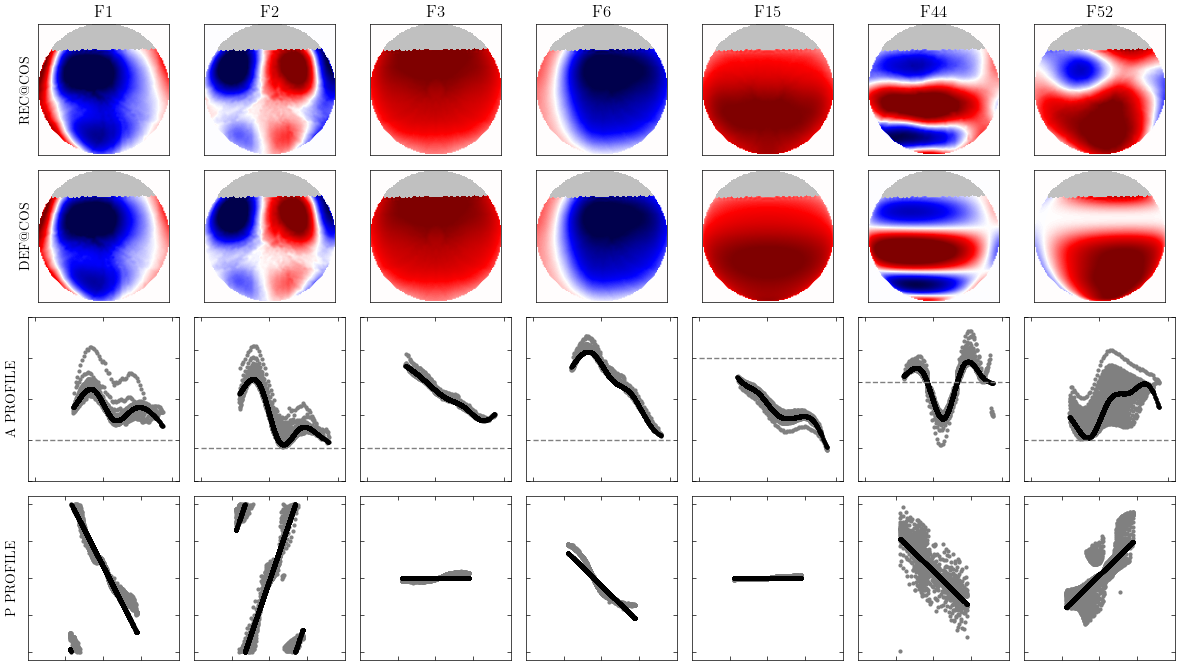}
    \caption{Result of Eclipse Mapping reconstruction. Each column contains maps and profiles for the frequency labelled at the top of the column. The top two rows contain the reconstructed maps as seen by the observer and the corresponding 'reference map' holding the "rosette" symmetry content of the former. The uniformly grey polar caps represent the uneclipsed regions, the integrated signals of which were fitted by additional virtual pixels. Row 3 (from top) shows the amplitude profile in function of the stellar co-latitude, with north on the left (0 degrees) and south to the right (180 degrees). Row 4 shows the phase profile in function of the stellar longitude, from -180 to +180 degrees. Longitude 0 at the centre corresponds to the stellar meridian. The individually computed values for the pixels are plotted with grey symbols, and the values interpolated from the fitted model to the same locations are plotted in black. The axis ticks are drawn every 90 degree, except for the vertical axis of the amplitude profiles. The tick labels have been omitted from the graphs to save space. The horizontal dashed lines in the amplitude graphs mark the individual zero levels.}
    \label{fig:em-maps-aligned}
\end{figure*}

\begin{figure*}
    \centering
    \includegraphics[width=\textwidth]{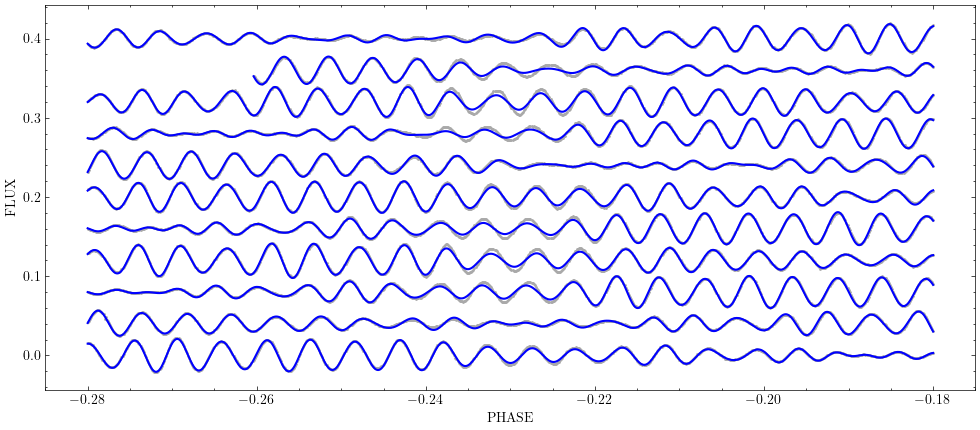}
    \caption{The fit of the combined Eclipse Mapping reconstructions to the data. Data for the individual orbital cycles are shown in separate rows from bottom to top. The input data are shown as grey circles, the fitted model as blue curves.}
    \label{fig:em-res}
\end{figure*}

\begin{figure*}
    \centering
    \includegraphics[width=\textwidth]{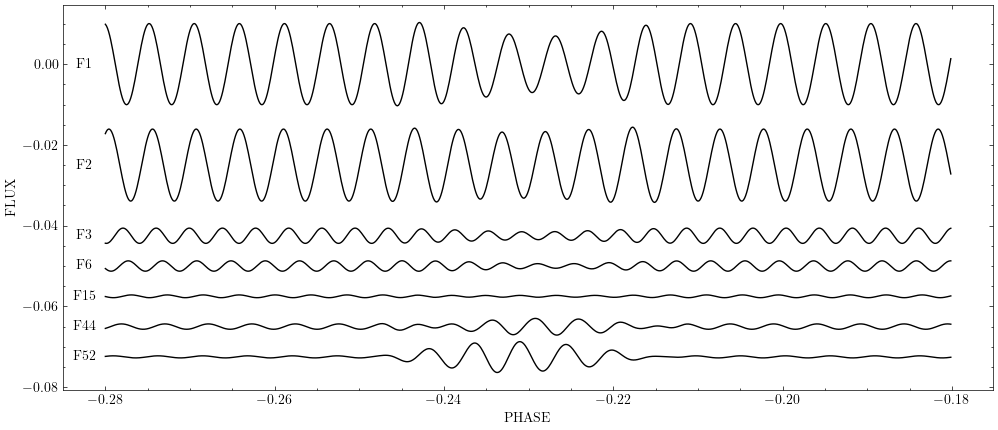}
    \caption{The individual flux contributions of the maps reconstructed by Eclipse Mapping. The curves, labelled by their frequencies, are shown on the same scale, only shifted vertically by arbitrary amounts.}
    \label{fig:em-grids}
\end{figure*}

The reconstructed maps and their amplitude and phase profiles are shown in
Figure~\ref{fig:em-maps-aligned}.
Contrary to our expectations, neither of the two strongest signals (F1, F2) emerged as radial mode. Both were mapped as sectoral modes: (2,-2) for F1 and (3,3) for F2, respectively. These mode numbers are unfortunately not recovered by the DF runs. We discuss the possible causes for all frequencies in the next section.

On the other hand, F3 and F15 were reconstructed as radial modes. DF runs also unambiguously attributed (0,0) to them.

F6 was reconstructed as a sectoral mode (1,-1). In comparison, DF failed to recover it, yielding equivalent modes instead: (0,0), (1,1), (2,0).
F44 and F52 were identified as (3,-1) and (2,1), respectively. These are typical hidden modes, showing the most pronounced modulations during the secondary eclipses. Because they are more distinguishable from the other modes in terms of their modulation pattern, DF procedures essentially gave the same results, but sometimes with opposite sign for $m$.

The fit to the data achieved by the EM reconstructions is shown in Figure~\ref{fig:em-res}. It shows the joint result of the two split, 5+2 runs. 
Figure~\ref{fig:em-grids} illustrates the contributions of the individual modes to the integrated flux for a single eclipse, as synthesised from the reconstructed maps.

\begin{table}[!h]
\centering
\caption{Results of Direct Fitting for the $\alpha=22^\circ$, and $\beta=22^\circ$ configuration. Notations are the same as in Tabe~\ref{tab:aligned-reconstructions}. The additional topmost row titled 'WF: predicted' lists the mode numbers inferred simultaneously with the axis direction angles derived from the Wigner coefficients of the best fitting multiplets.}
\label{tab:wigfit-df-mcmc}
\begin{tabular}{l|ccccc}
                & F1    & F2    & F3     & F6    & F15    \\
WF: predicted & (1,0) & (1,0) & (2,-1) & (2,0) & (2,1)  \\
\hline\hline
DF              & (0,0) & (0,0) & (0,0)  & (0,0) & (2,2)  \\
DFCLEAN         & (0,0) & (0,0) & (0,0)  & (0,0) & (2,2)  \\
MCMC            & (0,0) & (0,0) & (0,0)  & (2,0) & (2,-2) \\
percent         & 93.95 & 98.81 & 31.26  & 28.41 & 20.87  \\
odds            & 15.5  & 83.03 & 1.3    & 1.4   & 1.4    \\
odds of median  & 15.5  & 83.03 & 2.7    & 2.2   & 2.8   \\
\hline
\end{tabular}
\end{table}

\begin{table*}[!h]
\centering
\caption{Summary of the mode identification from the performed reconstruction for the selected frequencies.}
\label{tab:aligned-reconstructions}
\begin{tabular}{l|ccccccc}
               & F1     & F2    & F3    & F6     & F15    & F44    & F52    \\
               \hline\hline
EM             & (2,-2) & (3,3) & (0,0) & (1,-1) & (0,0)  & (3,-1) & (2,1)  \\
DF             & (0,0)  & (0,0) & (0,0) & (0,0)  & (1,-1) & (3,1)  & (2,-1) \\
DFCLEAN        & (0,0)  & (0,0) & (0,0) & (1.1)  & (1,-1) & (3,1)  & (2,-1) \\
MCMC           & (0,0)  & (0,0) & (0,0) & (2,0)  & (2,-2) & (3,-1) & (2,1)  \\
percent        & 92.5   & 99.4  & 27.8  & 17.7   & 13.2   & 18.1   & 25.0   \\
odds           & 12.4   & 168.5 & 1.2   & 1.7    & 1.1    & 1.01   & 1.05   \\
odds of median & 12.4   & 168.5 & 2.2   & 2.8    & 1.6    & 2.4    & 10.6  \\
\hline
\end{tabular}
\tablefoot{Mode identification results from Dynamic Eclipse Mapping (EM), Direct Fitting of $Y_\ell^m$ harmonics (DF), DF Cleaning and the {YLMCMC} algorithm. The last three rows are related to YLMCMC. "percent" is the percentage of the best candidate of the chain; "odds" is the ratio of probabilities of the best and second best candidates; and the "odds of median" is a similar ratio of the best candidate and the median of the rest of the sampled candidates.}
\end{table*}

\section{Discussion and conclusions}
\label{sec:discussion}

We invested a considerable amount of research in identifying the source components of the pulsations, given that the previous two works on this system \citep{Maceroni.etal.2014,Manzoori2020MNRAS.498.1871M} 
have arrived to conflicting conclusions in this regard. 

\citet{Maceroni.etal.2014} obtained the first binary model from Kepler photometry and extensive supplementary spectroscopy. They identified the high frequency p mode oscillations, containing most of the power, and also low frequency g mode oscillations. No evidence was found for tidal perturbations on the g modes. Similarly, faint signs on orbital modulation were found for some peaks, and no rotational splitting was detected. From the study of clustering of the p mode frequencies \citep{Breger.etal.2009}, they found it plausible that of the two strongest modes, F2 is a radial mode and F1 is a non-radial counterpart in the same cluster, provided that both pulsations originate on the secondary component. The common origin was supported by the behaviour of both the photometric and RV residuals from their EB analysis with respect to the orbital phase -- that the scatter intensifies during the secondary eclipse in both cases.

However, \citet{Manzoori2020MNRAS.498.1871M}, while performing an extensive tidal oscillation analysis on the system, concluded that F1 and F2 reside on different stars, on the grounds that they are very similar, as are also the stars (in mass, but not in radius, it should be emphasised, given that they have already evolved off the main sequence), therefore it is plausible that F1 and F2 are their radial mode frequencies. 
This conclusion appears to be supported by the values of the pulsation constant $Q = P_\text{puls} \, \left( \rho/\rho_\odot \right)$ \citep[][Table~II]{Breger.1979.puls.constant} for these frequencies as second and third overtones of the radial mode, based on the stellar masses and radii derived by \citet{Maceroni.etal.2014}.
As a further support, \citet{Manzoori2020MNRAS.498.1871M} presents results of the phase-modulation method of \cite{Murphy.etal.2014} applied to the data, confirming periodic time delay variations in opposite phase for the two frequencies.

Our results, however, do no support this scenario. Similar to \citet{Maceroni.etal.2014}, we find that the behaviour of the residuals from the EB analysis and after having subtracted all the identified pulsation frequencies as pure sinusoidals (Fig.~\ref{fig:kic-residuals}) shows without doubt that most of the eclipse modulations take place during the secondary eclipse.
There is clearly some modulation during the primary eclipse as well, indicating that some low-amplitude frequencies indeed do originate from the primary star. (We have identified one frequency, F10, as a possible mode of the primary.) Nevertheless, if the two {dominant} modes were on separate stars, the residuals would be amplified to comparable extents during the two eclipses of similar geometry (given that the orbit is almost sideways-on with $\omega \sim 22^\circ$). We do not see this in the data.
In addition, an eclipse mapping of the first 8 frequencies simultaneously on both stars attributed most of the modes to the secondary, leaving only F9 and F10 as of perhaps primary origin. (We note that F9 is approximately the double of F10.)
We also can't help noticing that the time delays of about $\pm 500$s of \citet[][Figure~B1]{Manzoori2020MNRAS.498.1871M} seem to be much larger than those expected from the binary orbit, which should vary between $116$s and $132$s depending on the orientation of the orbit, and it is about $\pm 120$s for the current configuration. The period of variation as assessed from the figure also seems to be too large compared to the orbital period. In any case, we have found that phase-modulation methods are very cumbersome to apply for multimode pulsations with many concurrent modes. Our attempt to apply it in the traditional way showed the delays of the two signals to vary synchronously rather than oppositely, but with an uncertain period. A folded version of the PM technique, using the known orbital period, only marginally reproduced the time delay curves expected for the current orbit, and the two delays still running together (Fig.~\ref{fig::pm-investigation}).

We now turn to the discussion of the mode identification results. A full pulsation analysis is beyond the scope of our paper, as it has already been covered in details by the preceding works \citep{Maceroni.etal.2014,Manzoori2020MNRAS.498.1871M}. Here we focus on assessing the feasibility of the various eclipse mapping methods regarding the mode identification. These methods use the surface sampling phenomenon of the mutual eclipses, depend on the orbital parameters of the binary system and basic stellar parameters (radii and limb darkening coefficients) only, therefore provide independent information for future asteroseismic investigations. 
We also did not attempt to map the very low amplitude g modes, as they would have challenged the methods beyond their capabilities.

The \'echelle diagram folded with the orbital frequency proved to be a crucial diagnostic tool in identifying the peaks modulated during the eclipses. Of the p modes, we successfully located 7 such candidates on the secondary star. They are however not the seven strongest amplitude pulsations. We possibly found one candidate (F10) on the primary too, but with less certainty.
A Direct Fitting method using multiplets was applied for the first five candidates to determine the orientation of the symmetry axis, and resulted in $\alpha \sim 22^\circ$ and $\beta \sim 22^\circ$ for the azimuthal and tilt angles, basically an aligned configuration. 
We then performed generic DF analyses in both tilted and aligned configurations for the same five candidates. Their results mostly agree, giving the same mode numbers, or equivalent mode numbers that are barely distinguishable in the present eclipse geometry. Therefore we assumed aligned configuration for the subsequent runs, and repeated the DF procedures for all the seven frequencies.

We also successfully reconstructed surface pulsation patterns with EM, and inferred best matching mode numbers. 
The two strongest modes were found to be sectoral, with (2,-2) and (3,3) for F1 and F2 according to EM. It came as a surprise that none of them seem to be radial.
In turn, F3 and F15 were found to be best approximated by radial modes in all mapping procedures involving various symmetry angle orientations. The pulsation constant for F3 is $Q=0.026$, which in theory can be attributed to the first overtone of the fundamental radial frequency of the secondary \citep[][Table~II]{Breger.1979.puls.constant} -- although, according to \citet{Maceroni.etal.2014}, both stars have evolved considerably from the main sequence, so it is not clear whether the classical pulsation constant can serve as a useful diagnostic in this case. As for F15, we note that it may be a combination frequency, as $F15 \approx 3 \cdot F12 - 3 \cdot F8$ (Table~\ref{tab:freqlist}), in which case its radial nature is no surprise.
F6 is best described by a retrograde sectoral mode (1,-1). F44 and F52 are two hidden modes that popped up apparently from nowhere during the \'echelle diagram analysis. Their initial ignorance hampered our first set of mappings, but after their discovered, not only did the analysis succeed, but they could also be identified as ($3,\pm 1$) and ($2,\pm 1$). 
We should mention however that the reconstructed pattern for the latter is less reliable. It can also be noted that all the surface patterns suffer from distortions arising from the interplay of the different frequencies, as a consequence of the small number of individual eclipses covered by the data compared to the number of mapped modes (11 eclipses for 7 frequencies).

In general the absolute value of $m$ could be uniquely determined, but not its sign. The results from EM have systematically opposite signs compared to the DF results. Figure~\ref{fig:synthetic-modulations} shows that, for the geometric configuration of KIC~3858884, modes with identical $\ell$ but oppositely signed $m$ are close to each other in terms of their modulations; therefore they can be mistaken for each other. The investigation of the associated side lobe structure may help to resolve the ambiguity (if the data quality allow it): the main peak is seen centered for $m=0$ and shifted to the left or right for $m<0$ and $m>0$, respectively.
A visual check of Figure~\ref{fig::echelle-modulations} shows that the main peak of F1 is located left to the maximum of the side lobe forest, while for F2 it is shifted to the right, supporting the values of $m$ found from EM rather than DF. Unfortunately this opportunity quickly fades with decreasing amplitudes. For F3 there is a marginal hint of centered location, i.e. $m=0$, which is indeed the value derived from all mappings; for F6 the peak seems left-shifted, pointing to a negative $m$. For F15 the structure is too sparse to draw a conclusion, and for the two 'hidden' modes the absence of additional side lobes 'islands' prevents a clear conclusion (cf. Figure~\ref{fig:synthetic-modulations}, lower panels). 

As for those leading pulsations that do not seem to be modulated at all, like F4, F5, F7, F8, F9, most of them are recognised as combination frequencies (cf.\ Table~\ref{tab:freqlist}), therefore they are not real peaks and are not expected to contribute to the modulations.

It is not clear why the DF and EM results disagree for F1 and F2 when they mostly match for the other modes, including the two low amplitude tesseral modes F44 and F52. Although the radial and the low-order sectoral modes give rise to similar modulations, one would expect that the strongest signals should be best differentiated. One reason may be that these modes differ the most from spherical harmonics, due to their larger amplitudes. The other source of disagreement could be that the direction of the pulsation axis as inferred from the photometric modulations is poorly determined. In our opinion this is the current bottleneck for further progress.

As a final conclusion, we successfully located seven pulsations on the secondary component, including the two strongest ones. Their modulations were detected in the frequency spectra of datasets with one or both types of eclipses being masked out. Crucial to this detection was the use \'echelle diagrams. We have also successfully identified these modes using Direct Fitting and Eclipse Mapping methods. We could uniquely determine the radial mode of the secondary as being F3. To our knowledge, for the first time ever, we found and identified hidden modes. 

We feel that our analysis is only scratching the surface of opportunities provided by eclipsing binaries with pulsating components and in particular this system, KIC~3858884, with its superb short-cadence \textit{Kepler} photometry data. 
As pointed out already by \citet{Maceroni.etal.2014}, a future spectroscopic survey of the system could be a tremendous help for mode identification. Apart from improving the atmospheric parameters and opening the way for spectroscopic mode identification methods, spectra of sufficient resolution and signal to noise ratio could also provide the possibility to fix the orientation of the symmetry axis -- currently the largest uncertainty factor in our photometry-based analysis -- using the Rossiter-McLaughlin effect.

\begin{acknowledgements}
We would like to thank László Molnár and Susmita Das for the discussions of possible asteroseismological implication regarding $\delta$ Scutis and for their modelling efforts.
This work was supported by the PRODEX Experiment Agreement No. 4000137122 between the ELTE E\"otv\"os L\'or\'and University and the European Space Agency (ESA-D/SCI-LE-2021-0025). 
This project has been supported by the LP2021-9 Lend\"ulet grant of the
Hungarian Academy of Sciences. 
This project has received funding from the HUN-REN Hungarian Research Network.
This paper includes data collected by the \textit{Kepler} mission. Funding
for the \textit{Kepler} mission is provided by the NASA Science Mission
directorate.
\end{acknowledgements}

%
%

\bibliographystyle{aa} 
\bibliography{BoA_ref}

\begin{appendix}
\section{Determined frequencies and additionally modellings}

The Table~\ref{tab:freqlist} lists the first 55 detected frequencies with the highest amplitude.

\begin{figure*}
    \centering
    \includegraphics[width=0.9\textwidth]{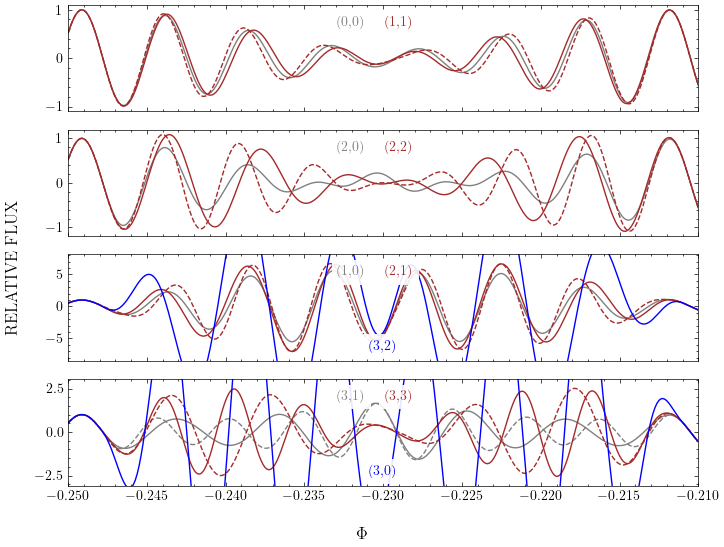}
    \caption{Synthetic modulations of considered ($\ell$,$m$) modes during the secondary eclipse. All modes were modelled with the same frequency and out of eclipse amplitude. The associated non-negative ($\ell$,$m$)-s are labelled with the proper colour in the middles of the panels. The dashed pairs of non-zero modes are the negative $m$ sibling of the same mode. They are grouped as a 'families' according their resemblance to each other.}
    \label{fig:synthetic-modulations}
\end{figure*}

\begin{figure*}
    \centering
    \includegraphics[width=0.9\textwidth]{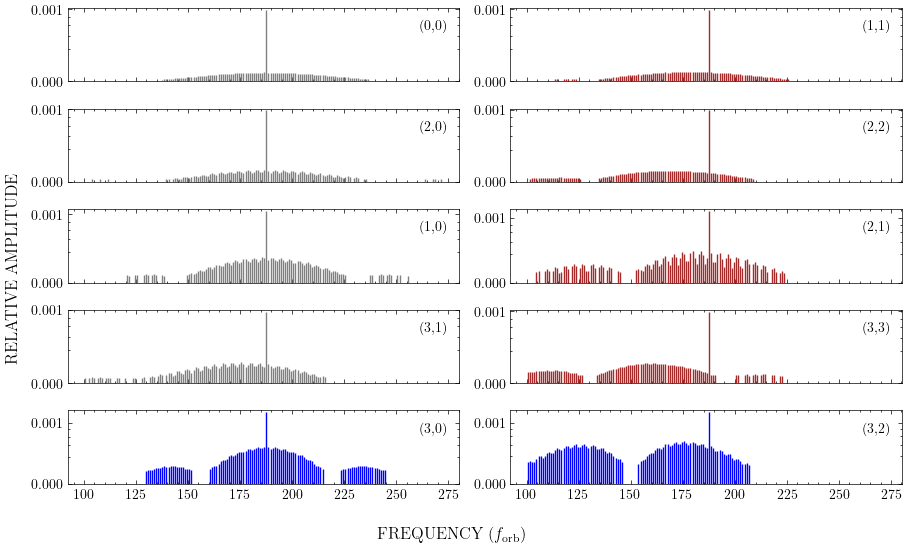}
    \caption{Fourier-spectra of all modelled, non-negative ($\ell$,$m$) modes calculated by \texttt{SigSpec}. The first four row presents all modes which are more distinguishable pairs from the different families shown in Fig.~\ref{fig:synthetic-modulations}. In the last row illustrates the hidden modes with the largest amplitude. The mode numbers are labelled in right top corner of each panel, the colours are meant to assistance to compare with the associated 'family member' in Fig.~\ref{fig:synthetic-modulations}. Note that the amplitudes are showed on a square-root scale for a better visibility of the side peak forests.}
    \label{fig:synthetic-spectra}
\end{figure*}

The synthesized lightcurves and corresponding Fourier spectra are presented in Fig.~\ref{fig:synthetic-modulations} and Fig.~\ref{fig:synthetic-spectra}, respectively.

\begin{table*}
    \centering
    \caption{First 55 detected frequencies with the highest amplitude.}
    \label{tab:freqlist}
    \begin{tabular}{lrrrrl}
    \hline
    id & f ($f_\text{orb}$) & f ($d^{-1}$) & amp & $\varphi$ (rad) & note \\
    \hline\hline
    F1 & 187.65 &  7.2245 & 0.009821 & 2.69 &  \\
    F2 & 193.95 &  7.4671 & 0.009052 & 4.87 &  \\
    F3 & 255.30 &  9.8292 & 0.001868 & 0.30 &  \\
    F4 & 194.96 &  7.5061 & 0.001801 & 2.45 &  $\text{F2}+f_\text{orb}$\\
    F5 & 174.81 &  6.7301 & 0.001550 & 2.72 &  \\
    F6 & 247.04 &  9.5111 & 0.001243 & 3.10 &  \\
    F7 & 381.60 & 14.6916 & 0.001182 & 0.88 &  $\text{F1}+\text{F2}$\\
    F8 & 304.31 & 11.7158 & 0.001009 & 5.56 &  \\
    F9 & 382.15 & 14.7128 & 0.000593 & 0.71 &  \\
    F10 & 191.08 &  7.3564 & 0.000539 & 3.59 &  \\
    F11 & 193.65 &  7.4555 & 0.000485 & 2.34 &  \\
    F12 & 382.80 & 14.7378 & 0.000458 & 6.13 &  \\
    F13 & 187.95 &  7.2361 & 0.000512 & 2.02 &  \\
    F14 &   1.00 &  0.0385 & 0.000432 & 2.60 & $f_\text{orb}$ \\
    F15 & 235.49 &  9.0665 & 0.000335 & 5.02 & $3\cdot \text{F12} -3\cdot \text{F8}$ \\
    F16 &  18.11 &  0.6972 & 0.000406 & 0.12 &  \\
    F17 & 188.59 &  7.2607 & 0.000348 & 2.30 &  \\
    F18 & 208.49 &  8.0268 & 0.000370 & 0.42 &  \\
    F19 & 253.46 &  9.7581 & 0.000364 & 4.72 &  \\
    F20 & 368.91 & 14.2031 & 0.000329 & 2.37 &  \\
    F21 & 298.89 & 11.5074 & 0.000317 & 3.96 &  \\
    F22 & 259.55 &  9.9926 & 0.000323 & 3.85 &  $3\cdot \text{F13} - \text{F8}$\\
    F23 &   6.30 &  0.2426 & 0.000291 & 2.14 &  $\text{F2}-\text{F1}$ \\
    F24 & 241.44 &  9.2956 & 0.000271 & 2.29 &  \\
    F25 & 186.42 &  7.1770 & 0.000258 & 4.14 &  \\
    F26 & 442.95 & 17.0537 & 0.000262 & 2.61 &  $\text{F1}+\text{F3}$\\
    F27 & 375.30 & 14.4491 & 0.000279 & 4.92 &  $2\cdot \text{F1}$\\
    F28 & 387.90 & 14.9341 & 0.000241 & 3.00 &  $2\cdot \text{F2}$\\
    F29 &  23.41 &  0.9013 & 0.000230 & 1.07 &  \\
    F30 & 268.39 & 10.3331 & 0.000238 & 5.66 &  \\
    F31 & 240.81 &  9.2712 & 0.000245 & 2.97 &  \\
    F32 & 359.01 & 13.8220 & 0.000220 & 0.82 &  \\
    F33 & 440.99 & 16.9781 & 0.000213 & 1.38 &  $\text{F2}+\text{F6}$\\
    F34 & 368.76 & 14.1972 & 0.000203 & 1.53 &  $\text{F2}+\text{F5}$\\
    F35 &  13.24 &  0.5097 & 0.000186 & 1.85 &  $4\cdot \text{F17}-3\cdot \text{F6}$\\
    F36 & 367.34 & 14.1426 & 0.000213 & 4.53 &  \\
    F37 & 180.72 &  6.9579 & 0.000221 & 1.62 &  \\
    F38 & 443.91 & 17.0904 & 0.000295 & 4.42 &  $\text{F3} + \text{F17}$\\
    F39 & 305.54 & 11.7633 & 0.000252 & 2.02 &  \\
    F40 & 382.67 & 14.7326 & 0.000173 & 5.84 &  \\
    F41 & 236.77 &  9.1156 & 0.000183 & 5.67 &  \\
    F42 & 388.91 & 14.9732 & 0.000174 & 0.68 &  $2\cdot \text{F2} + f_\text{orb}$\\
    F43 & 186.95 &  7.1976 & 0.000172 & 5.66 &  $\text{F13}-f_\text{orb}$\\
    F44 & 194.69 &  7.4956 & 0.000144 & 5.05 &  \\
    F45 & 293.12 & 11.2849 & 0.000159 & 5.44 &  \\
    F46 & 311.03 & 11.9745 & 0.000269 & 3.85 &  $3\cdot \text{F29}+ \text{F31}$\\
    F47 & 177.80 &  6.8452 & 0.000150 & 5.38 & $\text{F5} + 3\cdot f_\text{orb}$ \\
    F48 & 161.82 &  6.2301 & 0.000147 & 2.54 &  \\
    F49 &   3.04 &  0.1172 & 0.000148 & 0.64 &  \\
    F50 &  27.35 &  1.0531 & 0.000147 & 5.47 &  \\
    F51 & 311.02 & 11.9741 & 0.000226 & 5.38 &  \\
    F52 & 188.91 &  7.2729 & 0.000182 & 1.37 &  \\
    F53 & 296.52 & 11.4161 & 0.000142 & 1.55 &  \\
    F54 &  13.14 &  0.5057 & 0.000140 & 0.30 & $\text{F13}-\text{F5}$ \\
    F55 &  61.35 &  2.3621 & 0.000138 & 1.92 & $\text{F3}-\text{F2}$ \\
    \hline
    \end{tabular}
    \tablefoot{The associated frequencies in both units of orbital frequencies and c/d, the amplitude and phases are presented. In the last column the possible frequency combination is indicated as well.}
\end{table*}

\section{Probabilistic plots from YLMCMC}

\begin{figure*}[!h]
    \centering
    \includegraphics[width=0.9\textwidth]{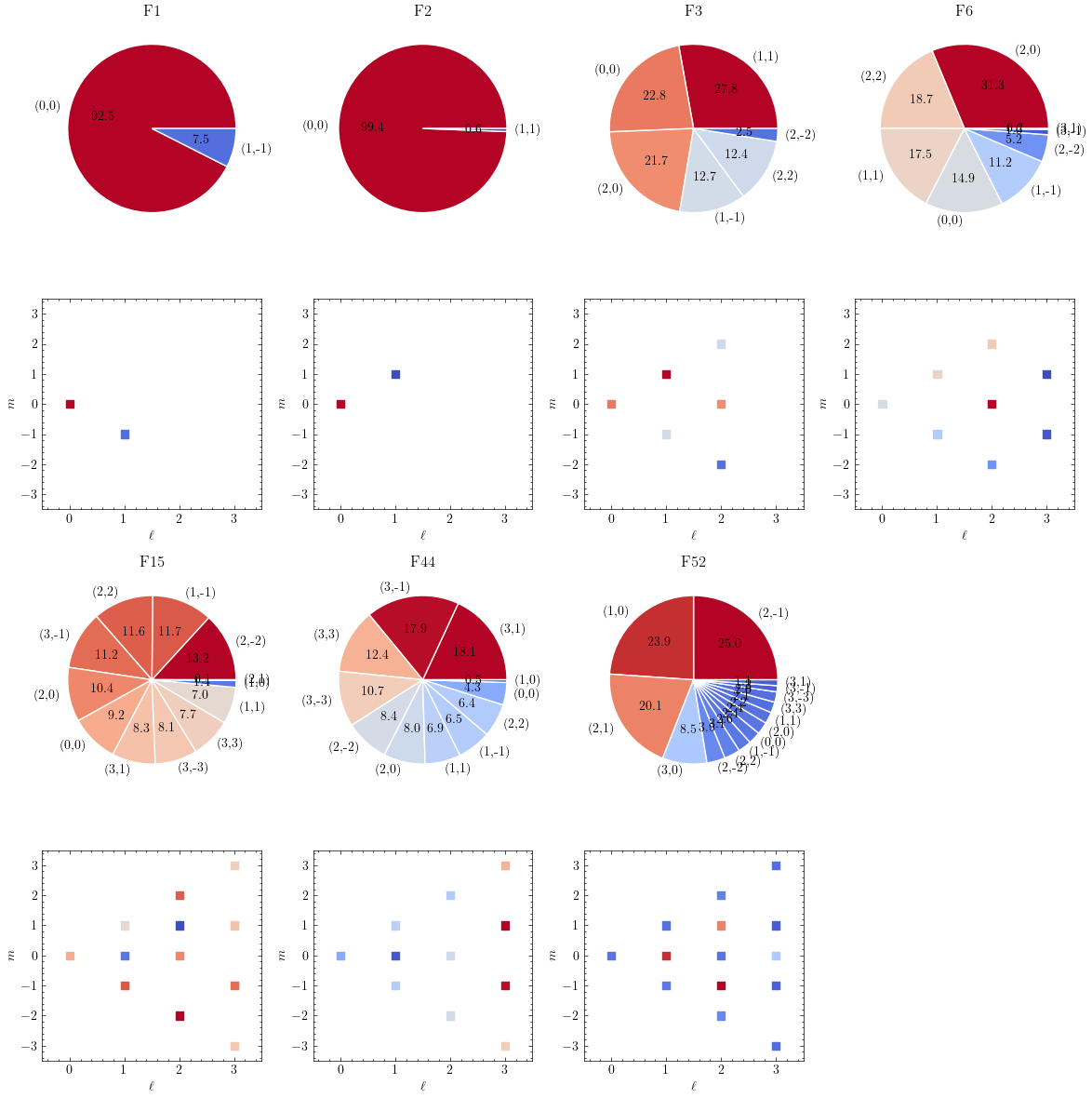}
    \caption{Result of YLMCMC, aligned configuration.}\label{fig:ylmcmc-aligned}
\end{figure*}

For completeness, Fig~\ref{fig:ylmcmc-aligned} is shown for illustration of all sampled ($\ell$,$m$) with information of the percentage.

\end{appendix}

\end{document}